\documentclass[aps,prx,twocolumn,superscriptaddress,notitlepage]{revtex4-2}
\usepackage{amsmath,amssymb,amsfonts}
\usepackage{graphicx, xcolor, physics}

\usepackage{caption}
\usepackage{subcaption}
\usepackage{ragged2e}

\DeclareCaptionJustification{myjust}{\justifying}

\usepackage{hyperref,todonotes, soul}

\usepackage{tikz}
\usepackage[most]{tcolorbox}
\usepackage{booktabs}
\usepackage{tabularx}
\usepackage{rotating}
\usepackage{rotating}
\usepackage{array}
\newcolumntype{L}[1]{>{\raggedright\arraybackslash}p{#1}}
\newcolumntype{C}[1]{>{\centering\arraybackslash}p{#1}}

\usepackage{eso-pic}
\usepackage{xcolor}

\begin{document}

\newcommand{\Red}[1]{\textcolor{red}{#1}}
\newcommand{\Dodo}[1]{\textcolor{blue}{#1}}
\newcommand{\SKG}[1]{\textcolor{purple}{#1}}
\newcommand{\chg}[1]{\textcolor{teal}{#1}}


\title{Generating GKP states using quantum dots inside a strongly coupled cavity}
\author{Viswatma Kamath}
\affiliation{Quanfluence Pvt. Ltd, Bengaluru, India}
\email{viswatmak@quanfluence.com}

\author{Ravi Mehta}
\affiliation{Quanfluence Pvt. Ltd, Bengaluru, India}

\author{Biman Chattopadhyay}
\affiliation{Quanfluence Pvt. Ltd, Bengaluru, India}

\author{Sandeep K.~Goyal}
\affiliation{Indian Institute of Science Education and Research, Mohali, India}

\begin{abstract}
GKP states enable fault-tolerant CV
quantum computation, but their generation remains experimentally challenging due to their highly non-Gaussian and infinite-energy ideal structure. In this work, we present a realistic and scalable protocol for generating finite-energy resource states, specifically the qunaught state,  using Schr\"odinger cat states generated
via a strongly coupled quantum dot-cavity system. Our scheme combines deterministic squeezed cat-state generation, cat-breeding protocols, and homodyne measurements. Using numerical simulations, we present a detailed analysis of the role of various parameters of the quantum-dot–cavity system in the generation of practical qunaught states. Furthermore, we quantify the trade-off between the fidelity and generation probability of these states and exploit the fact that accepting a structured set of homodyne outcomes can dramatically enhance the overall success rate. The proposed approach is compatible with integrated photonics and telecom wavelengths, offering a promising route toward scalable CV quantum information processing.
\end{abstract}

\maketitle
\section{Introduction}

Continuous Variable (CV) quantum information processing offers a powerful and flexible
framework for quantum communication, computation, and sensing, where information is
encoded in bosonic modes of light or collective excitations of harmonic oscillators \cite{Braunstein2005,Weedbrook2012,Serafini2017}.
Optical CV platforms are particularly attractive due to their low loss, natural compatibility with long-distance quantum networks, and the absence of stringent cryogenic requirements, which together enable practical scalability\cite{Braunstein2005,Weedbrook2012,Pirandola2020,Larsen2019}. However, it is
well known that Gaussian states and Gaussian operations alone are insufficient for
universal quantum computation or fault-tolerant quantum error correction
\cite{Lloyd1999,Braunstein2005,Weedbrook2012}.
This makes the generation of highly non-Gaussian resource states a core requirement for progress in CV quantum technologies.

Among the most prominent non-Gaussian resources are grid states, whose wavefunctions form a periodic lattice in phase space and which are stabilized by discrete displacement operators \cite{GKP2001}. The canonical example is the Gottesman–Kitaev–Preskill (GKP) state, which encodes a logical qubit into a single bosonic mode using a periodic grid structure in phase space \cite{GKP2001, Menicucci2014, glancy2006, Walshe2022}. The GKP encoding enables the correction of small displacement errors in both the position and momentum quadratures using only Gaussian operations and measurements, making it uniquely well suited for CV fault tolerance\cite{GKP2001, Menicucci2014, glancy2006, Campagne2020, Bourassa2021}

Unfortunately, ideal GKP states require infinite squeezing and energy, rendering them unphysical. Practical implementations therefore rely on approximate, finite-energy GKP states, whose performance is determined by their effective squeezing and the regularity of the grid structure. Within this framework, the qunaught state, denoted by $|Q_0\rangle$ , plays a central role as it enables the deterministic generation of GKP Bell states, thereby providing a crucial primitive for entanglement distribution and cluster-state construction in GKP-based architectures. The ability to generate high-fidelity approximate $|Q_0\rangle$ states under realistic constraints such as finite squeezing, loss, and imperfect operations is therefore a key requirement for scalable GKP-based quantum technologies.

Significant experimental progress toward GKP states has been achieved in trapped-ion
mechanical oscillators \cite{Fluhmann2019} and superconducting microwave cavities
\cite{Campagne2020}. These platforms allow deterministic preparation and error correction,
but face challenges related to scalability, fabrication complexity, and long-distance
connectivity. Optical implementations, by contrast, offer intrinsic compatibility with
quantum communication and photonic integration, but typically rely on probabilistic state
generation and suffer from low success rates
\cite{Vasconcelos2010,glancy2006}.

A particularly promising optical route to GKP states is based on Schrödinger cat states,
which are superpositions of coherent or squeezed coherent states. Cat states can be
progressively converted into grid-like states via \emph{cat-breeding} protocols, which
combine beam-splitter interference and conditional homodyne measurements
\cite{Lund2008,Ourjoumtsev2006,Ourjoumtsev2009,Weigand2019}. While conceptually simple and experimentally accessible, conventional cat breeding relies on strict post-selection—typically on a
single homodyne outcome—leading to exponentially suppressed success probabilities in
multi-round protocols and severely limiting scalability.

In parallel, advances in cavity quantum electrodynamics (QED) with semiconductor quantum
dots have enabled strong, coherent light–matter interactions at the single-photon level
\cite{Reiserer2015,Lodahl2015}. Quantum dots embedded in high-quality optical cavities
exhibit large cooperativity, fast optical nonlinearities, and spin-selective selection
rules that enable deterministic, state-dependent phase shifts upon photon reflection.
Importantly, recent developments show such systems can be engineered to operate at telecom wavelengths and are compatible
with on-chip photonic integration, making them attractive candidates for scalable optical quantum technologies.

In this work, we propose a realistic, end-to-end scheme for the high-rate generation of finite-energy grid resource states, combining deterministic cat-state generation in a quantum dot–cavity system with an optimized probabilistic breeding protocol.
First, we show how high-fidelity squeezed Schrödinger cat states can be deterministically generated via
spin-dependent reflection from a $\Lambda$-type quantum dot–cavity system. Second, we
employ a multi-round cat-breeding protocol to convert these states into grid-like
structures approaching the finite-energy qunaught state. Third, and crucially, we utilize the fact
that the success probability of cat breeding can be dramatically enhanced by accepting a
structured set of homodyne outcomes\cite{Pizzimenti25} that yield high fidelity, rather than post-selecting
only a single measurement result.

Using extensive numerical simulations, we quantify the trade-off between fidelity and
success probability and show that a two-round cat-breeding protocol can generate
high-quality qunaught states with experimentally feasible success rates. We further
analyze the impact of finite pulse duration, cavity loss, ground-state non-degeneracy, and
pure dephasing, and identify practical parameter regimes where these effects can be minimized. Our results establish quantum dot–cavity systems as a promising and scalable
platform for optical quantum computation using GKP states.

The remainder of this paper is organized as follows. In Section~II, we review the current state of the art in grid-state generation with particular emphasis on optical and industry-relevant approaches, and position the present work within this landscape. In Section~\ref{sec:non_gaussian_resources}, we introduce the non-Gaussian resource states and related concepts central to this work, including the cat-breeding protocol and the dependence of the qunaught state quality on input squeezing, and its teleportation fidelity. In Section~\ref{sec:theory}, we describe the light--matter interaction formalism for a traveling optical pulse interacting with a cavity-coupled quantum emitter, and show how a $\Lambda$-type quantum dot operated in the strong coupling regime enables deterministic, state-dependent phase shifts upon reflection, and demonstrate how this interaction can be exploited to generate high-fidelity squeezed Schr\"odinger cat states. Appendix~\ref{sec:io_theory} presents the framework for simulating the of cavity-atom interaction, in particular the cascaded-system formalism and the impact of realistic imperfections such as finite pulse duration, cavity loss.
Section~\ref{sec:sim_results} presents numerical simulation results,
beginning with the dependence of qunaught state fidelity and teleportation
performance on the input squeezing parameter $r$, which determine
the optimal cat-state parameters $(\alpha, r)$. We then simulate the
cavity dynamics using these parameters and analyse the quality of
the generated cat states under realistic conditions. Finally, we quantify
the trade-off between fidelity and success probability in the two-round
qunaught generation protocol.

Section~\ref{sec:comparison_gkp_cluster} compares the present protocol with representative state-of-the-art experimental GKP architectures. Finally, Section~\ref{sec:conclusion} summarizes the main conclusions of this work and outlines possible directions for future extensions, including higher-round breeding protocols.

\section{State of the Art and Industry Relevance of GKP State Generation}

To date, the most advanced realizations of GKP states have been achieved in microwave-frequency platforms, including superconducting circuits and trapped-ion systems~\cite{Fluhmann2019,Campagne2020}. These systems benefit from strong intrinsic nonlinearities and precise control over bosonic modes, enabling deterministic preparation and repeated quantum error correction of grid states. While these demonstrations represent important milestones, their reliance on cryogenic environments or localized architectures limits their direct applicability to scalable and networked quantum technologies.

In contrast, optical and photonic implementations of GKP states offer several compelling advantages. Photonic systems operate naturally at room temperature, are inherently compatible with low-loss fiber networks, and support high-bandwidth quantum information processing. These features make them uniquely suited for large-scale and distributed quantum computing, as well as quantum communication applications.

The primary challenge in optical GKP state generation lies in the need for
non-Gaussian resources. While strong optical nonlinearities could, in principle, provide such
resources, most current approaches instead rely on measurement-induced
non-Gaussianity, where non-classical features are generated probabilistically
through conditional measurements. This led to the development of cat-breeding protocols in which interference of smaller
non-Gaussian states produces increasingly structured phase-space lattices~\cite{glancy2006, Vasconcelos2010}.

 Advances in the generation of high-quality squeezed states, optical cat states, and near-unity-efficiency homodyne detection have enabled higher fidelities. More recent proposals incorporate hybrid approaches, combining optical modes with matter-based nonlinearities to enhance performance and move toward near-deterministic generation~\cite{Hastrup2022}. Unlike bulk-optical or photon-subtraction-based schemes, the use of a solid-state,
chip-integrated emitter enables deterministic generation of high-fidelity squeezed
cat states at rates set by the cavity bandwidth. Moreover, operation at optical or
telecom wavelengths makes the protocol naturally compatible with fiber-based quantum
communication and large-scale photonic integration. These features position
quantum dot–cavity platforms as a promising bridge between high-quality bosonic
encodings and industry-relevant quantum technologies.

A major conceptual development in the photonic paradigm is the integration of GKP states into CV cluster-state architectures for measurement-based quantum computing~\cite{Menicucci2014}. In this framework, large-scale entangled cluster states are generated deterministically using Gaussian operations, which are readily scalable via time- and frequency-multiplexing in optical systems. By embedding GKP-encoded qubits into CV cluster states, one obtains a hybrid architecture that combines the scalability of Gaussian photonics with the error-correcting capabilities of discrete-variable encoding. In this setting, logical operations are implemented through adaptive homodyne measurements and classical feedforward, while GKP states provide protection against small displacement errors and enable fault-tolerant thresholds. This architecture forms the foundation of scalable photonic quantum computing efforts, including those pursued in integrated photonics platforms~\cite{Bourassa2021}. 

Table~\ref{tab:gkp_soa} summarises the key figures of merit of
these representative schemes alongside the present work, highlighting
the complementary strengths of each approach in terms of fidelity, success rate, and scalability.

\begin{table*}[t]
\centering
\includegraphics[width=1\textwidth]{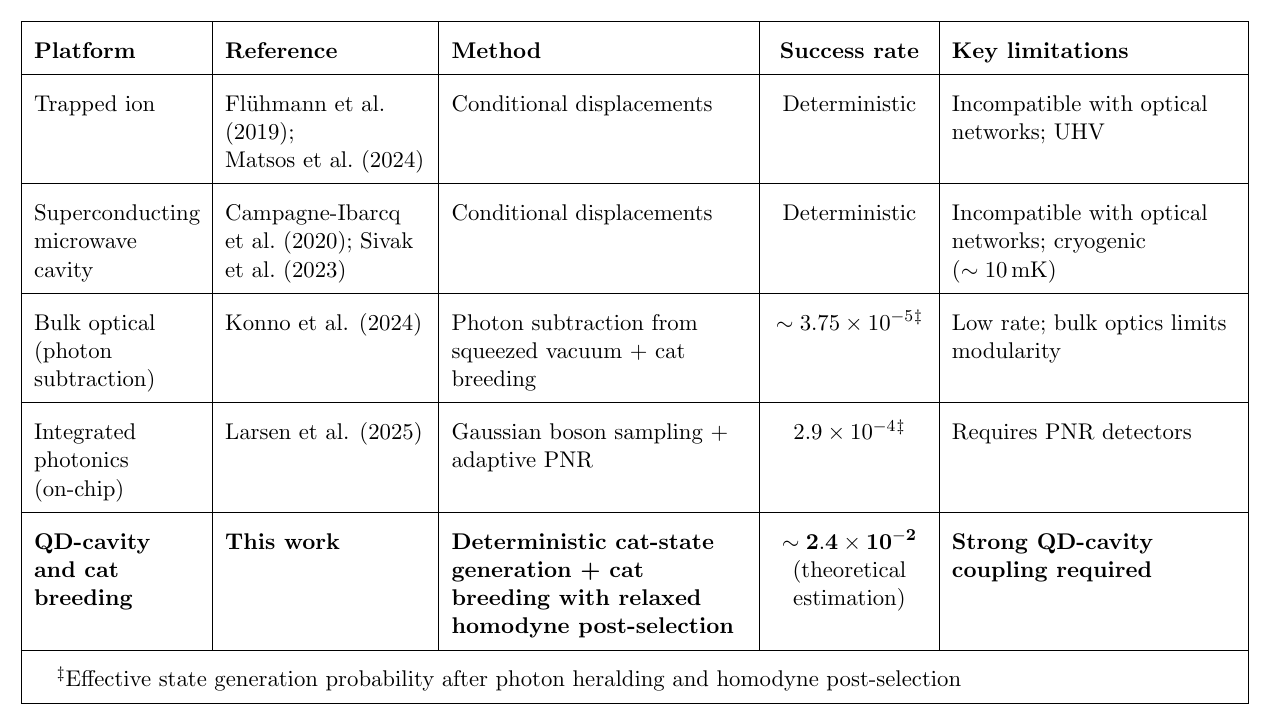}
\caption{State-of-the-art experimental demonstrations for GKP state generation. Industry relevance qualitatively assesses
compatibility with scalable fabrication, optical networking, and room-temperature operation.\\
\textit{Note: This work is theoretical; it is intended as a proposal motivating a concrete experimental roadmap toward realization of the protocol.}
}
\label{tab:gkp_soa}
\end{table*}

\section{Non-Gaussian Resources for Continuous-Variable Quantum Information}
\label{sec:non_gaussian_resources}
While Gaussian states and Gaussian operations form a powerful and experimentally
accessible toolbox, they are insufficient for universal computation or
fault-tolerant encoding \cite{Lloyd1999,Braunstein2005,Weedbrook2012}.
In this section, we introduce the non-Gaussian states central to this work:
Schrödinger cat states, GKP grid states, qunaught states,  finite-energy
approximations thereof.

Further, we discuss the protocols that exploit these resource states:
the cat-breeding protocol\cite{Glancy2008}, homodyne detection; a relaxed post-selection strategy\cite{Pizzimenti25} that substantially
improves the success probability of each breeding round; and the use of
$|Q_0\rangle$ states as an entangled resource for CV quantum teleportation,
which serves as an operational benchmark for the quality of the generated
states.

\subsection{Schrödinger Cat States}
\label{subsec:cat_states}

\captionsetup{
  justification=justified,
  singlelinecheck=false
}

Schrödinger cat states are superpositions of squeezed coherent states with macroscopically distinct phase-space displacements and constitute a paradigmatic class of non-Gaussian optical states~\cite{Sanders1989, Glancy2008, gilchrist2004, Ourjoumtsev2006}. For a single bosonic mode, a squeezed coherent state
$|\alpha, \xi\rangle$ is defined as - 
\begin{align}
    \ket{\alpha, \xi} = \hat{D}(\alpha) \hat{S}(\xi) \ket{0},
\end{align}
where $\hat{D}(\alpha) = \exp(\alpha \hat{a}^\dagger - \alpha^* \hat{a})$ is the displacement operator and $\hat{S}(\xi)$ is the squeezing operator.
\begin{align}
    \hat{S}(\xi) = \exp\left[ \frac{1}{2} \left( \xi^* \hat{a}^2 - \xi \hat{a}^{\dagger 2} \right) \right], \quad \xi = r e^{i\theta}.
\end{align}
 Here $a$ is the corresponding annihilation operator of the optical mode, $\alpha$ is the displacement amplitude, $r$ is the squeezing amplitude and $\theta$ the squeezing angle.
The squeezing amplitude $r$, can be represented in decibels by $\mathrm{dB} = -10\log_{10}(e^{-2r})$.
 The squeezed cat states take the form:
\begin{align}
    \ket{\mathrm{SC}_\pm(\alpha, \xi)}
    &= \frac{1}{\mathcal{M}_\pm}
        \left( \ket{\alpha, \xi} \pm \ket{-\alpha, \xi} \right), \label{eq:SCsqueezed}
\end{align}
where $\mathcal{M}_\pm$ normalizes the states~\cite{Sanders1989, Glancy2008, gilchrist2004, Ourjoumtsev2006}.

\begin{center}
  \begin{figure}
\captionsetup{justification=myjust, singlelinecheck=false}

  \includegraphics[width=\columnwidth]{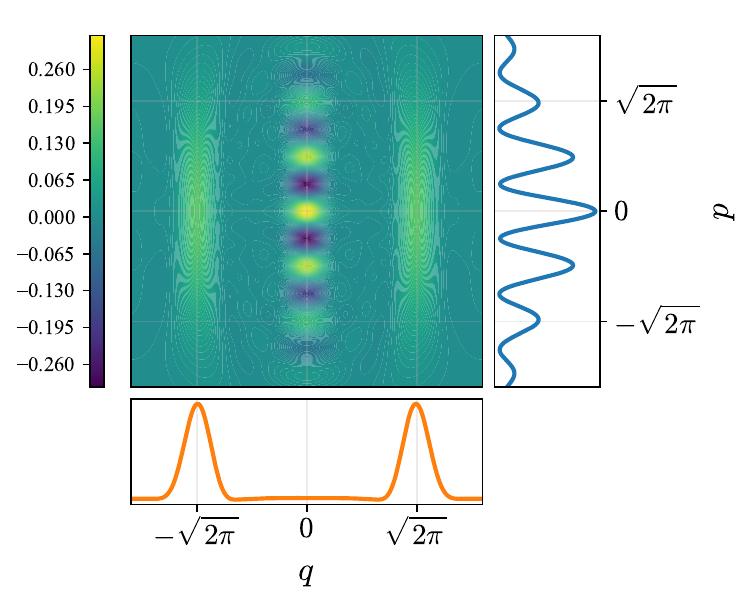}
  \caption{Wigner distribution of $\ket{\mathrm{SC}_\pm(\alpha, \xi)}$, along with the $\hat{q}$ and $\hat{p}$ marginals. Here $\alpha = \sqrt{\pi}$ is chosen as the optimal displacement amplitude for the two-round cat-breeding protocol (see Section~\ref{subsec:squeezing_Q0}), while the squeezing $\xi = 0.98$ ($r = 0.98$, corresponding to $\approx 8.5~\text{dB}$) is selected to illustrate a representative ideal squeezed cat state. The justification for these values chosen for our protocol are given in Sec~\ref{sec:sim_results}. In the position quadrature, the Wigner function exhibits two well-separated Gaussian lobes, while the momentum marginal carries pronounced interference fringes that reflect the quantum coherence between the two components. The Wigner distribution takes negative values at certain phase-space points, confirming the non-Gaussian nature of the state.}
  \label{fig:Wig-cat}
\end{figure}
\end{center}

Cat states exhibit interference fringes and negativity in their Wigner functions (see Fig.~\ref{fig:Wig-cat}),
and have been experimentally realized using photon subtraction, cavity QED, and
superconducting circuits \cite{Ourjoumtsev2007,Hacker2019,Ofek2016}.
For many quantum-information applications, squeezed cat states are employed as squeezing enhances the phase-space structure of cat states and significantly
improves their suitability as precursors for grid-state generation
\cite{Lund2008}.

\subsection{Gottesman-Kitaev-Preskill States}
\label{subsec:gkp_states}

GKP states encode a logical qubit into a single
bosonic mode using a periodic lattice structure in phase space
\cite{GKP2001,Grimsmo2021,Brady2024,LatticePersp2022}.
The ideal logical basis states are defined as
\begin{align}
|0_L\rangle &\propto \sum_{n\in\mathbb{Z}} |q = 2n\sqrt{\pi}\rangle, \\
|1_L\rangle &\propto \sum_{n\in\mathbb{Z}} |q = (2n+1)\sqrt{\pi}\rangle,
\end{align}
where $|q\rangle$ denotes an eigenstate of the position quadrature
$\hat{q}=(\hat{a}+\hat{a}^\dagger)/\sqrt{2}$.
In the canonical square-lattice encoding, the GKP position-space wavefunction forms
a Dirac comb with fundamental spacing $2\sqrt{\pi}$ between adjacent peaks,
with the logical states $|0_L\rangle$ and $|1_L\rangle$. In phase space, this forms a discrete lattice structure which is why such states are also called grid states. 

GKP states are uniquely powerful because small displacement errors can be corrected
using only Gaussian operations and measurements, enabling fault-tolerant CV quantum
computation \cite{Menicucci2014,glancy2006}.
However, ideal GKP states require infinite squeezing and energy, rendering them 
unphysical.
\subsection{Finite-Energy GKP States and the \texorpdfstring{$|Q_0\rangle$}{Q0} State}
\label{subsec:qnaught}
Practical implementations therefore rely on finite-energy GKP states, in which the
delta-function peaks of the ideal states are replaced by Gaussians of width $\Delta_q$ and multiplied
by an overall Gaussian envelope of width $1/\Delta_p$ \cite{Menicucci2014, Royer2020,glancy2006,Albert2018}. The expressions for the finite-energy logical states $|0_L\rangle$ and $|1_L\rangle$ in the position basis, $\psi_0(q) = \langle q | 0_L \rangle$ and $\psi_1(q) = \langle q | 1_L \rangle$, read:
\begin{align}
    \psi_0(q) &\propto \sum_{s \in \mathbb{Z}} \exp\left(-\frac{(q - 2 s \sqrt{\pi})^2}{2 \Delta_q^2}\right) \exp\left(-\frac{(\Delta_p 2s\sqrt{\pi})^2}{2}\right) , \\
    \psi_1(q) &\propto \sum_{s \in \mathbb{Z}} \exp\left(-\frac{(q - (2 s + 1)\sqrt{\pi})^2}{2 \Delta_q^2}\right) \exp\left(-\frac{(\Delta_p 2s\sqrt{\pi})^2}{2}\right) .
\end{align}
These states are normalizable, physical, and can be generated  using experimentally feasible displacement and squeezing operations~\cite{Menicucci2014}.

Of particular importance is the qunaught state $|Q_0\rangle$, a symmetric
finite-energy grid state with lattice spacing $\sqrt{2\pi}$ given by - 
\begin{align} 
    \psi_{Q0}(q) &\propto \sum_{s \in \mathbb{Z}} \exp\left(-\frac{(q - s \sqrt{2\pi})^2}{2 \Delta_q^2}\right) \exp\left(-\frac{(\Delta_p s\sqrt{2\pi})^2}{2}\right) 
    \label{eq:Q0_wavefunction}
\end{align}
The qunaught state plays a central role in measurement-based CV quantum
computation, as interference of two identical $|Q_0\rangle$ states on a beam
splitter generates entangled GKP Bell pairs \cite{Menicucci2014,Weigand2019}.

\subsection{Cat Breeding}
\label{subsec:cat_breeding}

Cat breeding is a probabilistic protocol that converts Schrödinger cat states into
states with increasingly pronounced grid-like structure in phase space.
The protocol consists of interfering two identical cat states on a balanced beam
splitter, followed by homodyne measurement of one output mode
\cite{Ourjoumtsev2007,Lund2008}.

\begin{figure}[h]
\captionsetup{justification=myjust, singlelinecheck=false}

  \includegraphics[width=0.35\textwidth]{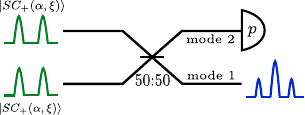}
  \caption{Cat breeding setup. Two identical squeezed cat states $|\mathrm{SC}_+(\alpha,r)\rangle$ are interfered on a balanced (50:50) beam splitter. A homodyne measurement of the $p$-quadrature on mode~2 then projects mode~1 onto a three-peaked grid-like state (see text). Two of these output grid states can be interfered again to produce states with 5 peaks and so on.}
  
  
  \label{Fig:Cat-Breeding setup using intereference and quadrature measurement on mode 2.}
\end{figure}

The protocol begins with two identical squeezed cat states of amplitude $\alpha$ and squeezing parameter $r$ (see Fig.~\ref{Fig:Cat-Breeding setup using intereference and quadrature measurement on mode 2.}):
\begin{equation}
    \ket{\mathrm{SC}_+(\alpha,r)}
    \approx \frac{1}{\sqrt{2}} \left( |\alpha,r\rangle + |-\alpha,r\rangle \right).
\end{equation}
Two such cat states are interfered on a balanced beam splitter, whose action on annihilation operators of the two Bosonic modes $a_1$ and $a_2$ is
\begin{equation}
    \hat{a}_1 \rightarrow \frac{1}{\sqrt{2}}(\hat{a}_1 + \hat{a}_2), 
    \qquad
    \hat{a}_2 \rightarrow \frac{1}{\sqrt{2}}(\hat{a}_1 - \hat{a}_2).
\end{equation}
Using the displacement addition rule,
\begin{equation}
    |\alpha,r\rangle_1 |\beta,r\rangle_2 
    \rightarrow 
    \left|\frac{\alpha+\beta}{\sqrt{2}},r\right\rangle 
    \left|\frac{\alpha-\beta}{\sqrt{2}},r\right\rangle,
\end{equation}
the joint input state
\begin{equation}
    |\Psi_{\mathrm{in}}\rangle
    = |\mathrm{SC}_+(\alpha,r)\rangle
      \otimes
      |\mathrm{SC}_+(\alpha,r)\rangle
\end{equation}
becomes, after the beam splitter,
\begin{align}
    |\Psi_{\mathrm{BS}}\rangle
    &= \frac{1}{2} \Big(
         \left|\sqrt{2}\alpha,r\right\rangle |0,r\rangle
         + |0,r\rangle \left|\sqrt{2}\alpha,r\right\rangle  \nonumber \\
    &\qquad
         + \left|-\sqrt{2}\alpha,r\right\rangle |0,r\rangle
         + |0,r\rangle \left|-\sqrt{2}\alpha, r\right\rangle
      \Big).
\end{align}

A homodyne measurement of the $p$-quadrature on mode~$2$,
with outcome $p\approx 0$, projects the state onto
\begin{equation}
    |\Psi_{\mathrm{out}}\rangle
    \propto
    \langle p=0 | \Psi_{\mathrm{BS}}\rangle.
\end{equation}
Since $\ip{p=0}{\alpha, r} = \mathrm{constant}$ for all real $\alpha$ and $r$, the state of mode~$1$ collapses to 
\begin{equation}
    |\Psi_{\mathrm{out}}\rangle
    \propto
    2|0,r \rangle
    + |\sqrt{2}\alpha, r\rangle
    + |-\sqrt{2}\alpha, r\rangle.
\end{equation}
For real $\alpha$, $r$, the three-peaked structure approximates a discrete grid in the phase space.

The cat-breeding protocol can be applied recursively: the output of one breeding round serves as the input to the next, each time producing a state with a more pronounced grid-like structure in phase space. After n rounds, the resulting state has $2^n+1$ prominent peaks, progressively approximating a finite-energy grid state. In principle, the two input states to each round need not be identical; however, using identical inputs simplifies the protocol considerably by reducing the number of free parameters and making the scheme easier to implement experimentally. We therefore adopt this symmetric configuration throughout.
Two important parameters govern the quality of the output state. First, the displacement amplitude $\alpha$ must be chosen carefully — its optimal value depends on both the number of breeding rounds and the target grid state. For the two-round protocol targeting the qunaught state, the optimal choice is $\alpha = \sqrt{\pi}$ (see Section~\ref{subsec:squeezing_Q0}). Second, the success probability of each breeding round is not fixed but depends on how strictly one post-selects on the homodyne measurement outcome; a more relaxed acceptance criterion trades a small reduction in fidelity for a substantially higher success rate, as discussed in detail in Section~\ref{sec:improving_success}.


Cat-breeding protocols provide a scalable pathway from experimentally accessible
cat states to approximate grid states, and form the basis of several optical GKP
state-generation proposals \cite{Vasconcelos2010,Weigand2019}.

\subsection{Improving the success rate in the Cat-breeding protocol}
\label{sec:improving_success}
\begin{figure*}[t]
\captionsetup{justification=myjust, singlelinecheck=false}

\centering
\begin{subfigure}[b]{\columnwidth}
\includegraphics[width=1\linewidth]{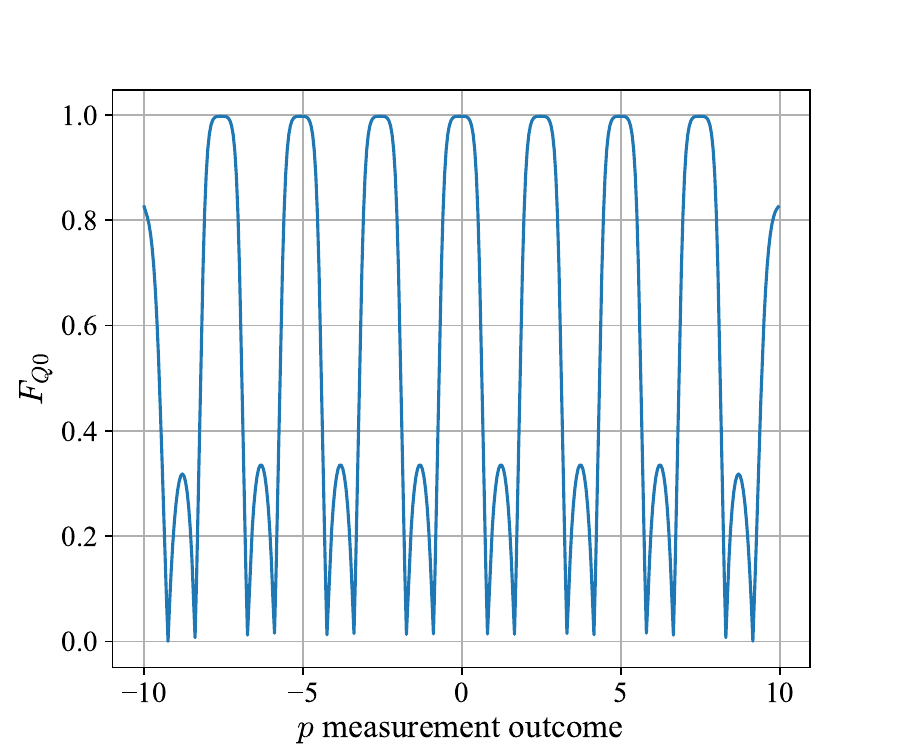}
  \caption{}
  \label{fig:3pfid}
\end{subfigure}
\hfill
\begin{subfigure}[b]{\columnwidth}
    \includegraphics[width=\linewidth]{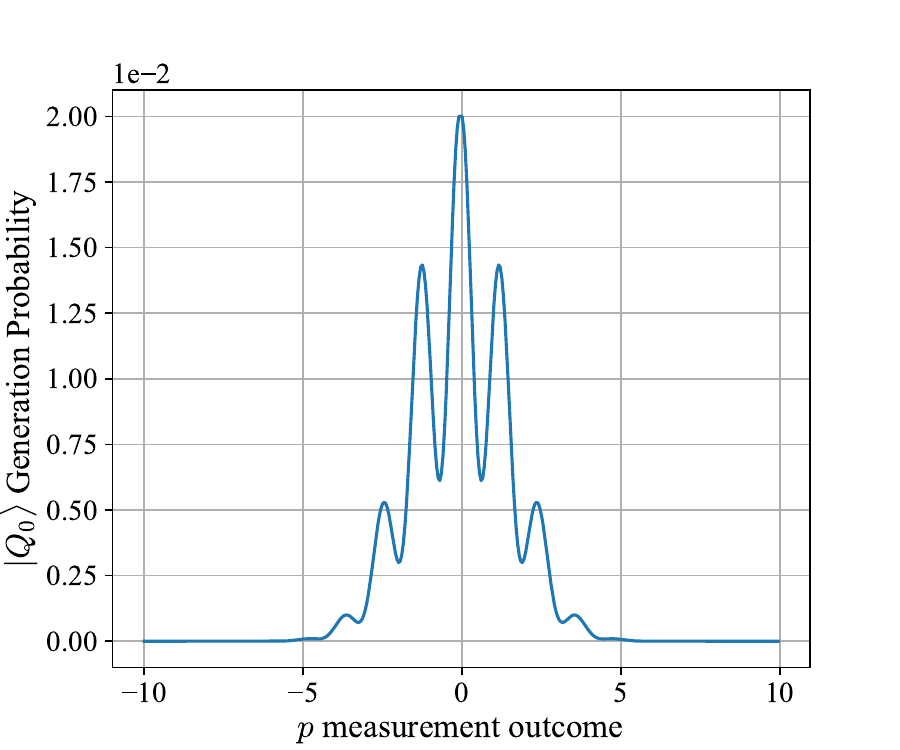} 
    \caption{}
  \label{fig:3pp}  
\end{subfigure}

  \caption{Subfig.~\ref{fig:3pfid} shows $p$ outcome vs.\ fidelity $F_{Q_0}(p)$ for a single round of cat breeding. The periodic high-fidelity windows can be used to relax measurement post-selection conditions, substantially enhancing the success rate of the desired state. Subfig.~\ref{fig:3pp} shows the probability of $|Q_0\rangle$ generation for the corresponding $p$ outcome.}
  \label{fig:p_fidelity}
\end{figure*}

In the standard cat-breeding protocol, the generation of higher-quality non-Gaussian states such as finite-energy qunaught states—is typically conditioned on a homodyne measurement outcome $p=0$ in one of the output modes. While this strict post-selection criterion yields high-fidelity states, it comes at the cost of a low success probability, since the probability density at exactly $p=0$ is small for realistic finite-energy states.

A more efficient strategy is to relax this condition and accept a broader range of homodyne outcomes $p$, provided that the resulting conditional state maintains a fidelity above a chosen threshold with respect to the target state \cite{Pizzimenti25}. The dependence of the conditional-state fidelity on the homodyne measurement outcome $p$ exhibits a characteristic oscillatory structure, as shown in Fig.~\ref{fig:3pfid}. High-fidelity regions are not confined to $p=0$ but occur periodically at multiple values of $p$.

Formally, let $F(p)$ denote the fidelity between the conditional output state $\rho(p)$ and the target state $\ket{Q_0}$,
\begin{equation}
    F(p) = \bra{ Q_0}\rho_p \ket{ Q_0}.
\end{equation}
Instead of post-selecting only $p=0$, we define an acceptance window
\begin{equation}
    \mathcal{P}_{\mathrm{acc}} = \{\, p \; | \; F(p) \geq F_{\mathrm{th}} \,\},
\end{equation}
where $F_{\mathrm{th}}$ is a chosen threshold fidelity (for example, $F_{\mathrm{th}} = 0.9$). All homodyne outcomes within $\mathcal{P}_{\mathrm{acc}}$ are retained, while outcomes outside this set are discarded.

The overall success probability of the breeding step is then given by
\begin{equation}
    P_{\mathrm{succ}}
    = \int_{\mathcal{P}_{\mathrm{acc}}} \! dp \; P(p),
\end{equation}
where $P(p)$ is the probability density of obtaining outcome $p$, illustrated in Fig.~\ref{fig:3pp}. Although $P(p)$ is peaked around $p=0$ and decays for larger $|p|$, the presence of multiple high-fidelity regions at nonzero $p$ allows for a substantial increase in $P_{\mathrm{succ}}$ when these outcomes are included.

This multi-outcome post-selection approach is particularly advantageous in multi-step breeding protocols, where the overall success probability would otherwise decrease exponentially with the number of breeding rounds. By retaining all measurement outcomes that yield sufficiently high fidelity, the protocol becomes substantially more resource-efficient and experimentally viable as shown in Fig.~\ref{fig:p_fidelity}.

\subsection{GKP teleportation using qunaught states}
\label{sec:teleportation}
Beyond state fidelity, a key operational benchmark for $|Q_0\rangle$ states is their
performance in quantum teleportation protocols. For finite-energy GKP states, the
teleportation fidelity depends on the effective squeezing of the grid peaks and
provides a direct measure of error-correction capability. 

Using the CV teleportation protocol \cite{Braunstein1998}  we evaluate
the teleportation fidelity of logical states encoded in the GKP basis.

We consider three modes: the input mode (mode~0) carrying the state $|\psi_{\rm in}\rangle$ to be teleported, and two ancilla modes (modes~1 and~2) initialized in $|Q_0\rangle$ states. The entangled resource is prepared by interfering the two $|Q_0\rangle$ ancilla states on a $50{:}50$ beam splitter $\hat{B}_{12}$,
\begin{equation}
|\Phi\rangle_{12} =
\hat{B}_{12} |Q_0\rangle_1 \otimes |Q_0\rangle_2 .
\end{equation}
This produces an entangled two-mode state (a GKP-encoded analogue of a Bell state); writing out the explicit wavefunction in the position basis gives a superposition of correlated grid peaks. The joint three-mode state is then - 
\begin{equation}
|\Psi_{012}\rangle =
|\psi_{\rm in}\rangle_0 \otimes |\Phi\rangle_{12}.
\end{equation}
The input mode~0 is then mixed with ancilla mode~1 via a beam splitter $\hat{B}_{01}$, after which Bell measurements are performed in the CV sense: a $\hat{q}$ homodyne measurement on mode~0 and a $\hat{p}$ homodyne measurement on mode~1, where $\hat{q}$ and $\hat{p}$ are the beam-splitter-transformed quadratures. The conditional state of mode~$2$ for measurement outcomes $(q,p)$ is

\begin{equation}
|\psi_{2}(q,p)\rangle =
\frac{\langle p|_1 \langle q|_0 |\Psi_{012}\rangle}
{\sqrt{\mathbb{P}(q,p)}},
\end{equation}
with the joint probability of obtaining outcomes $(q,p)$ given by

\begin{equation}
\mathbb{P}(q,p) =
\big|
\langle p|_1 \langle q|_0 |\Psi_{012}\rangle
\big|^2 .
\end{equation}
The receiver must then apply a logical Pauli correction to recover the
teleported state. For each measurement outcome $(q,p)$ we therefore
consider the set of logical Pauli operations
$\{\hat{I}, \hat{X}_L, \hat{Y}_L, \hat{Z}_L\}$ and define the
outcome-specific teleportation fidelity as

\begin{equation}
F(q,p) =
\max_{\hat{\sigma} \in
\{\hat{I},\hat{X}_L,\hat{Y}_L,\hat{Z}_L\}}
\Big|
\langle \psi_{\rm in} |
\hat{\sigma} |\psi_{2}(q,p)\rangle
\Big|^2 .
\end{equation}
The teleportation fidelity for a given input state is obtained by
averaging over all measurement outcomes weighted by their probabilities,

\begin{equation}
\bar{F}_{\mathrm{tel}} =
\int dq \, dp \;
\mathbb{P}(q,p)\, F(q,p).
\end{equation}
Finally, to characterize the teleportation performance over the logical
qubit space, we average over a representative set of logical states
\[
|\psi_{\rm in}^{(i)}\rangle \in
\{\ket{0_L}, \ket{1_L},
\ket{+_L}, \ket{-_L},
\ket{+i_L}, \ket{-i_L}\},
\]
giving the overall teleportation fidelity

\begin{equation}
F_{\mathrm{tel}} =
\frac{1}{N_{\rm states}}
\sum_{i=1}^{N_{\rm states}}
\bar{F}_{\mathrm{tel}}^{(i)} .
\end{equation}


\section{Light-matter interaction}
\label{sec:theory}

This section develops the theoretical framework for light–matter interactions between a traveling optical pulse and a cavity-coupled quantum emitter. We begin in Section~\ref{subsec:cavity_qed_model} by introducing the cavity QED description of a two-level emitter interacting with a single-sided optical cavity using input–output theory. In Section~\ref{subsec:lambda_system}, we extend this model to a $\Lambda$-type system, where state-dependent coupling gives rise to conditional phase shifts of the reflected optical field. Finally, in Section~\ref{subsec:cat_state}, we show how this mechanism enables the deterministic generation of non-Gaussian photonic states, specifically squeezed Schr\"odinger cat states, via spin–photon entanglement and projective measurement.

\subsection{Cavity QED Model for a Localized Quantum Emitter}
\label{subsec:cavity_qed_model}
We describe the basic cavity QED
interaction of a  two-level emitter in an optical cavity interacting with light. We consider a single-sided optical cavity interacting with a localized quantum emitter, such as an atom or a semiconductor quantum dot. 

The Hamiltonian of the coupled emitter–cavity system, within the rotating-wave
approximation, is described by the Jaynes–Cummings model
\begin{equation}
\hat{H}_{\mathrm{JC}}
=
\hbar \omega_c \hat{a}^\dagger \hat{a}
+
\hbar \omega_0 \hat{\sigma}_{\mathrm{ee}}
-
\hbar g \left(
\hat{a}\hat{\sigma}_{\mathrm{eg}}
+
\hat{a}^\dagger \hat{\sigma}_{\mathrm{ge}}
\right),
\label{eq:JC}
\end{equation}
where $\omega_c$ denotes the cavity resonance frequency, $\omega_0$ the
transition frequency of the two-level emitter, and $g$ the coherent
emitter–cavity coupling strength. The operators $\hat{a}$ and
$\hat{a}^\dagger$ annihilate and create a photon in the cavity mode
respectively, while $\hat{\sigma}_{ij} = |i\rangle\langle j|$ are the
atomic transition operators between the ground state $|g\rangle$
and excited state $|e\rangle$.

The cavity couples to a one-dimensional continuum of external modes through a
partially transmitting mirror, leading to photon loss at rate $\kappa$. In
addition, the emitter undergoes spontaneous emission into non-cavity modes at
rate $\gamma$. These dissipative processes are incorporated using the standard
input–output formalism \cite{Gardiner1985,Collett1984}.

The Heisenberg–Langevin equations for the cavity field and emitter coherence are
\begin{align}
\dot{\hat{a}} &=
-\left(i\Delta_c + \frac{\kappa}{2}\right)\hat{a}
- g \hat{\sigma}_{\mathrm{ge}}
- \sqrt{\kappa}\,\hat{a}_{\mathrm{in}}(t),
\label{eq:a_eom}
\\
\dot{\hat{\sigma}}_{\mathrm{ge}} &=
-\left(i\Delta_0 + \frac{\gamma}{2}\right)\hat{\sigma}_{\mathrm{ge}}
+ g \hat{a}\left(\hat{\sigma}_{\mathrm{gg}} - \hat{\sigma}_{\mathrm{ee}}\right),
\label{eq:sigma_eom}
\end{align}
where $\Delta_c = \omega_c - \omega_p$ and $\Delta_0 = \omega_0 - \omega_p$ denote
detunings from the central frequency $\omega_p$ of the incident pulse.

The output field operator is related to the intracavity and input fields by
\begin{equation}
\hat{a}_{\mathrm{out}}(t)
=
\hat{a}_{\mathrm{in}}(t)
+
\sqrt{\kappa}\,\hat{a}(t).
\label{eq:input_output}
\end{equation}

In the weak-excitation limit, $\langle\sigma_{\mathrm{gg}}\rangle \approx 1$ and $\langle\sigma_{\mathrm{ee}}\rangle \approx 0$, and under steady-state conditions these equations yield a linear reflection coefficient
\begin{equation}
r(\omega_p) \equiv
\frac{a_{\mathrm{out}}}{a_{\mathrm{in}}}
=
\frac{[i\Delta_c-\kappa/2][i\Delta_0+\gamma/2]+g^2}
{[i\Delta_c+\kappa/2][i\Delta_0+\gamma/2]+g^2}.
\label{eq:reflection_general}
\end{equation}

In the case of  resonant driving, $\omega_p=\omega_c=\omega_0$,  Eq.~\eqref{eq:reflection_general} simplifies to
\begin{equation}
r =
\frac{-\kappa\gamma/4 + g^2}{\kappa\gamma/4 + g^2}.
\label{eq:reflection_resonant}
\end{equation}

Two limiting cases are of particular importance:
\begin{enumerate}
\item \emph{Uncoupled emitter} ($g=0$): the reflection coefficient is
\begin{equation}
r = -1,
\end{equation}
corresponding to a $\pi$ phase shift acquired by the reflected field.
\item \emph{Strong coupling} ($g^2 \gg \kappa\gamma$): the reflection coefficient approaches
\begin{equation}
r \approx +1,
\end{equation}
and the photon is reflected without acquiring a phase shift.
\end{enumerate}
Thus, the presence or absence of coupling between the emitter and the cavity directly determines whether the reflected field acquires a relative phase of $0$ or $\pi$. This binary phase response is the fundamental mechanism enabling conditional phase gates.

\begin{figure}
\captionsetup{justification=myjust, singlelinecheck=false}
  \includegraphics[width=\columnwidth]{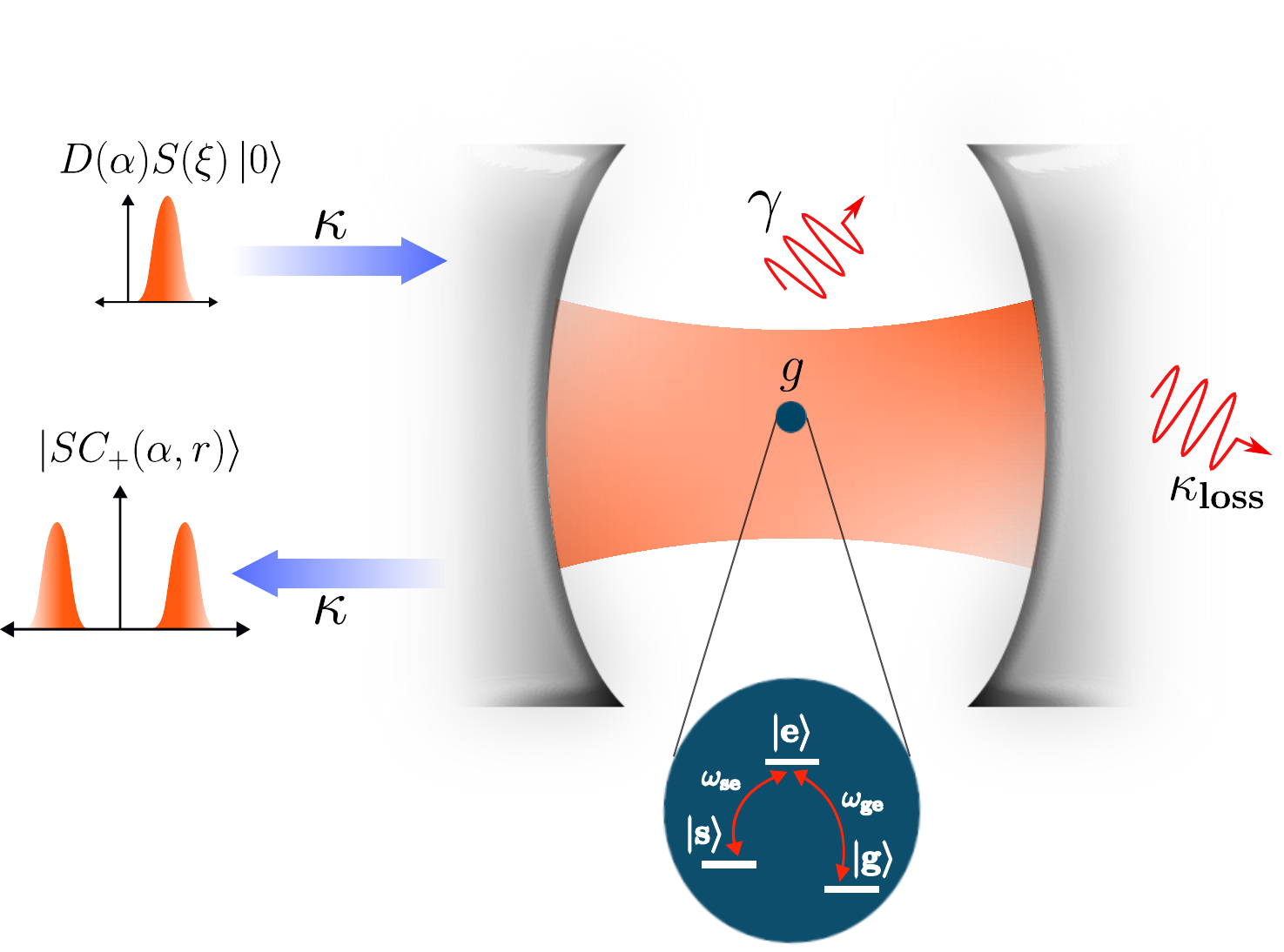}
  \centering
  \caption{Generating cat states using cavity QED: A quantum dot with ground states $\ket{\mathrm{g}}$ and $\ket{\mathrm{s}}$ is
embedded in a strongly coupled optical cavity. Only the
$\ket{\mathrm{s}} \leftrightarrow \ket{\mathrm{e}}$ transition is resonant
with the cavity mode; the $\ket{\mathrm{g}} \leftrightarrow \ket{\mathrm{e}}$
transition remains dark. The spin is initialized in an equal superposition $(|\mathrm{g}\rangle + |\mathrm{s}\rangle)/\sqrt{2}$, and a displaced squeezed optical pulse $|\alpha,r\rangle$ is incident on the cavity.
Upon reflection, the light acquires a spin-dependent phase shift: the coupled ($|\mathrm{s}\rangle$) transition leads to near-zero phase, while the uncoupled ($|\mathrm{g}\rangle$) transition results in a $\pi$ phase shift. This produces an entangled spin--photon state $(|\mathrm{s}\rangle|\alpha,r\rangle + |\mathrm{g}\rangle|-\alpha,r\rangle)/\sqrt{2}$. A subsequent Hadamard rotation on the spin followed by a projective measurement in the $\{|\mathrm{s}\rangle,|\mathrm{g}\rangle\}$ basis collapses the optical field into an even or odd squeezed cat state $|\mathrm{SC}_{\pm}(\alpha,r)\rangle$.}
  \label{fig:atom-cavity}
\end{figure}

\subsection{State-Dependent Interaction in a $\Lambda$ System}
\label{subsec:lambda_system}

To realize state-dependent optical nonlinearities, we consider a $\Lambda$-type
emitter with two long-lived ground states $|\mathrm{g}\rangle$ and $|\mathrm{s}\rangle$ coupled to a
common excited state $|\mathrm{e}\rangle$ (see Fig.~\ref{fig:atom-cavity}). Such a $\Lambda$-system can be physically realised using atoms, trapped ions, or semiconductor quantum dots.
For our protocol, we consider a quantum dot embedded in a high-finesse single mode optical cavity, interacting with a light pulse.
The $\Lambda$-system is placed inside a single-mode optical cavity which couples to only one of the transitions (say $\ket{\mathrm{s}}\leftrightarrow \ket{\mathrm{e}}$).
In this case, the interaction Hamiltonian takes the form
\begin{equation}
\hat{H}_{\mathrm{int}}
=
\hbar g
\left(
\hat{a}\,|\mathrm{e}\rangle\langle s|
+
\hat{a}^\dagger\,|\mathrm{s}\rangle\langle e|
\right),
\label{eq:lambda_H}
\end{equation}
while the state $|\mathrm{g}\rangle$ remains effectively uncoupled. As a result, the
reflection coefficient of an incident pulse depends on the internal state of the
emitter.

On resonance and in the strong-coupling regime $g^2 \gg \kappa\gamma$, the reflection
amplitude approaches $+1$ when the emitter is coupled and $-1$ when it is not. An
incident coherent or squeezed light therefore undergoes the transformation
\begin{equation}
|\mathrm{g}\rangle|\alpha, r\rangle \rightarrow |\mathrm{g}\rangle|-\alpha,r\rangle,
\qquad
|\mathrm{s}\rangle|\alpha,r\rangle \rightarrow |\mathrm{s}\rangle|\alpha,r\rangle,
\label{eq:conditional_phase}
\end{equation}
realizing a controlled $\pi$ phase shift between the emitter and the optical field.

Preparing the emitter in a superposition of ground states generates
spin–photon entanglement, which forms the basis for the deterministic generation
of non-Gaussian photonic states discussed in the following sections.

\subsection{Generating Schr\"odinger cat states}
\label{subsec:cat_state}
The squeezed cat states can be generated by leveraging spin-dependent light–matter interactions in a $\Lambda$-type emitter coupled to an optical cavity.
The system is initialized by preparing the quantum dot spin in a coherent superposition
\begin{equation}
    |\psi_{\mathrm{QD}}\rangle = \frac{1}{\sqrt{2}} \left( |\mathrm{g}\rangle + |\mathrm{s}\rangle \right).
\end{equation}
A displaced squeezed optical pulse $|\alpha,r\rangle$ is then incident on the cavity. The coupling of the atomic (or quantum dot) transition determines the phase acquired by the optical pulse upon reflection. 
Due to the strong coupling between the cavity and the $|\mathrm{s}\rangle \leftrightarrow |\mathrm{e}\rangle$ transition, the reflection coefficient acquires zero phase whereas the zero coupling between the cavity and the  $|\mathrm{g}\rangle \leftrightarrow |\mathrm{e}\rangle$ results in a $\pi$ phase to the reflected pulse. 
As a result, the joint spin–photon state after reflection becomes entangled:
\begin{equation}
    |\Psi_{\mathrm{ent}}\rangle
    = \frac{1}{\sqrt{2}} \left( |\mathrm{s}\rangle |\alpha,r\rangle + |\mathrm{g}\rangle |-\alpha,r\rangle \right).
\end{equation}
This state represents a Schr\"odinger cat–like entangled state between a discrete-variable spin qubit and a CV optical field.

To convert this entanglement into a purely photonic cat state, a unitary rotation is applied to the quantum dot spin. Specifically, performing a Hadamard operation, $\mathcal{H}$ in the $\{|\mathrm{s}\rangle, |\mathrm{g}\rangle\}$ basis,
\begin{equation}
    \mathcal{H} |\mathrm{s}\rangle = \frac{1}{\sqrt{2}} (|\mathrm{s}\rangle + |\mathrm{g}\rangle), 
    \qquad
    \mathcal{H} |\mathrm{g}\rangle = \frac{1}{\sqrt{2}} (|\mathrm{s}\rangle - |\mathrm{g}\rangle),
\end{equation}
transforms the joint state into
\begin{align}
    \mathcal{H} |\Psi_{\mathrm{ent}}\rangle
    = \frac{1}{2} \Big[
        |\mathrm{s}\rangle (|\alpha,r\rangle + |-\alpha,r\rangle)
        + |\mathrm{g}\rangle (|\alpha,r\rangle - |-\alpha,r\rangle)
    \Big].
\end{align}

Finally, a projective measurement of the quantum dot spin in the $\{|\mathrm{s}\rangle, |\mathrm{g}\rangle\}$ basis collapses the optical field onto a Schr\"odinger cat state. Conditioning on the measurement outcome yields either an even or odd cat state,
\begin{equation}
    |SC_{\pm}(\alpha,r)\rangle
    = \frac{1}{\sqrt{2}} \left( |\alpha,r\rangle \pm |-\alpha,r\rangle \right).
\end{equation}

This protocol describes an idealized scenario in which the spin-dependent
phase shift is perfectly implemented and the optical pulse acquires a uniform
phase upon reflection. In this limit, the generation of high-fidelity squeezed
cat states follows directly from coherent light–matter interaction and
projective measurement.
In practice, however, several non-ideal effects can degrade the quality of the
generated states. These include finite cooperativity, cavity losses, emitter
decay into undesired modes, and distortions arising from the finite bandwidth
of the input pulse. 
In the next subsection, we incorporate these effects, including a full quantum treatment of the pulse dynamics, and analyse the impact of such imperfections on the quality of the generated cat states.

\section{Simulation results}
\label{sec:sim_results}

\subsection{Dependence of \texorpdfstring{$|Q_0\rangle$}{Q0} State Quality on Input Squeezing}
\label{subsec:squeezing_Q0}

Squeezing controls both the sharpness of the grid peaks and the width of the Gaussian envelope.

After $n$ rounds of cat breeding, the output state approaches a grid state with $2^n +1$ number of prominent peaks. Consequently, the amount of available  squeezing limits the
maximum number of breeding rounds that can be usefully applied, as additional rounds
fail to generate new well-resolved peaks.

\begin{figure}[t]
\captionsetup{justification=myjust, singlelinecheck=false}

  \centering
  \includegraphics[width=\columnwidth]{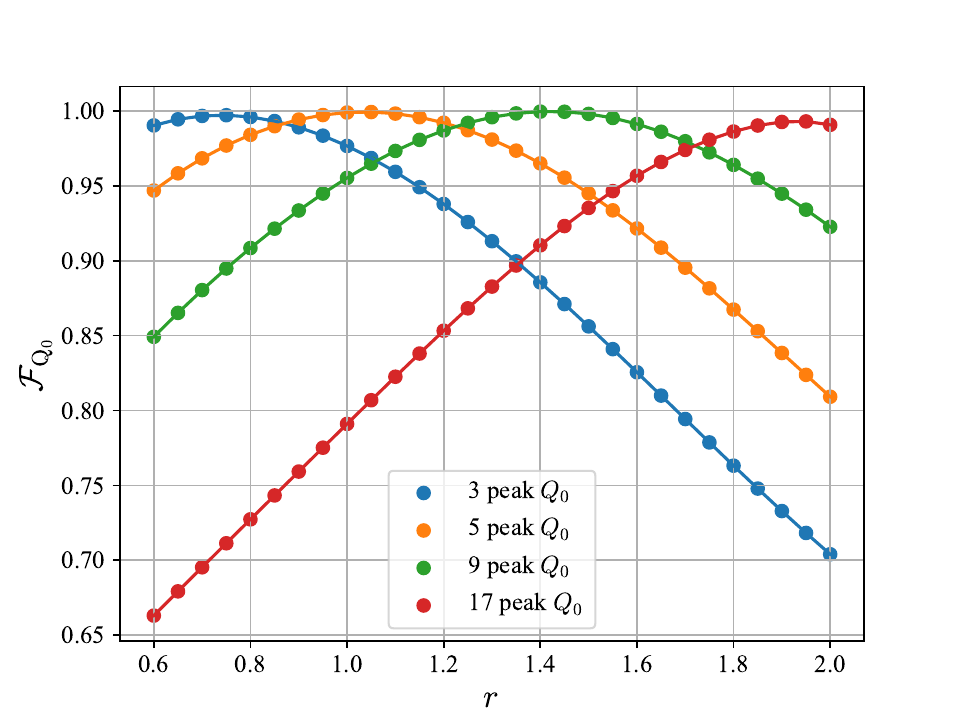}
  \caption{Fidelity of the generated $|Q_0\rangle$ state as a function of the input
  squeezing parameter $r$ for different numbers of cat-breeding rounds. Increasing
  the number of breeding rounds relaxes the squeezing requirement for achieving a
  given target fidelity.}
  \label{fig:fidelity_vs_r}
\end{figure}

To quantify this effect, we compute the fidelity
\begin{equation}
\mathcal{F}_{Q_0}(r,n)
=
\bra{Q_0^{\mathrm{ideal}}}
\rho_{Q_0}^{(n)}(r)
\ket{Q_0^{\mathrm{ideal}}},
\end{equation}
where $\rho_{Q_0}^{(n)}(r)$ denotes the numerically generated state after $n$ cat-breeding
rounds starting from an initial squeezed cat state of squeezing $r$. Figure~\ref{fig:fidelity_vs_r} shows the
resulting fidelity as a function of $r$ for one-, two-, three-, and four-fold cat
breeding.

Figure~\ref{fig:fidelity_vs_r} reveals two key trends. First, for a given number of breeding rounds, the fidelity rises as squeezing increases, reaches a maximum, and then decreases. The fidelity peaks at an optimal squeezing $r_{\rm opt}$ because the Gaussian envelope weights of the bred state are fixed by the beam-splitter interference and match the squeezing-dependent envelope
weights of the ideal finite-energy $|Q_0\rangle$ (Eq.~\eqref{eq:Q0_wavefunction})
only at a single value of $r=r_{\rm opt}$, for the three-peak state. For $r < r_{\rm opt}$ the outer peaks are
over-weighted relative to the ideal state, while for $r > r_{\rm opt}$
they are under-weighted, in both cases reducing the overlap with the target. Second, the squeezing value at which the maximum fidelity is achieved increases with the number of breeding rounds. This is because each additional round roughly doubles the number of grid peaks and therefore requires sharper individual peaks and hence higher squeezing -- to resolve them correctly.

\subsection{Dependence of \texorpdfstring{$\mathcal{F}_{\mathrm{tel}}$}{Ftel} on input squeezing}
\label{subsec:telep_num}
\begin{figure}[t]
  \centering
\captionsetup{justification=myjust, singlelinecheck=false}

  \includegraphics[width=\columnwidth]{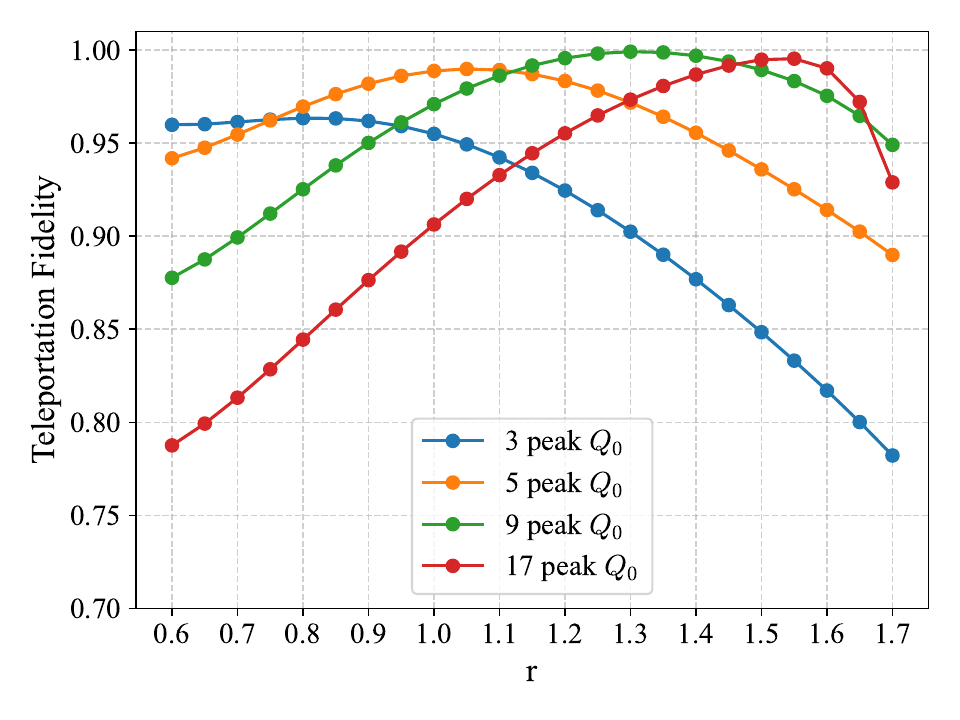}
  \caption{Teleportation fidelity using the generated $|Q_0\rangle$ states as a
  function of input squeezing $r$ for different cat-breeding depths. Improved
  squeezing and additional breeding rounds enhance the teleportation performance,
  approaching the fault-tolerant threshold for GKP-based protocols.}
  \label{fig:tele_vs_r}
\end{figure}

Following from the discussions in Sec.~\ref{sec:teleportation}., Fig.~\ref{fig:tele_vs_r}
shows the teleportation fidelity, $\bar{F}_{\mathrm{tel}}$ as a function of $r$ for the corresponding
$|Q_0\rangle$ states.

These results also confirm the trend of the $\ket{Q_0}$ state fidelity. We also see that for a target teleportation fidelity additional breeding rounds increase the minimum required squeezing. This, along with results in Sec.~\ref{subsec:squeezing_Q0} show that, for a fixed target state fidelity (both $F_{Q_0}$ and $F_{tel}$), increasing the number of breeding rounds raises the required squeezing. This insight is useful for experimental design. In particular, if the available squeezed-light source is limited to \( r = 0.7 \), it is optimal to use only a single round of cat breeding to obtain the best \( \ket{Q_0} \) state; additional breeding rounds would in fact degrade the resulting state. Taking into account both the state fidelity (Figure~\ref{fig:fidelity_vs_r}) and the teleportation fidelity (Figure~\ref{fig:tele_vs_r}), we adopt a two-round breeding protocol with $\alpha = \sqrt{\pi}$ and $r = 0.98$ ($\approx 8.5~\mathrm{dB}$), which yields the best trade-off between achievable fidelity and experimental feasibility. These are the cat-state parameters used for simulating the cavity-QD interaction seen in the following subsection and impact the realistic implementations of both qunaught and the GKP
state generation and teleportation-based CV quantum computation. 
\cite{GKP2001,Menicucci2014,Albert2018}. 

\subsection{Practical considerations for QD-cavity modelling}
\label{sec:cqed_practical}
With the optimal cat-state parameters $(\alpha, r)$ established, we now
turn to the feasibility of generating such states in a realistic quantum
dot--cavity platform.
The cavity is designed such that it has a single mode resonance at 1550 nm. The quantum dot placement ensures that the mode density is strongest at its location. 

The quantum dot is modeled as a three-level \(\Lambda\)-system with two stable ground states $\ket{\mathrm{g}}$ and $\ket{\mathrm{s}}$, and one excited state $\ket{\mathrm{e}}$. The cavity mode couples only to the \(|\mathrm{s}\rangle \leftrightarrow |\mathrm{e}\rangle\) transition, while \(|\mathrm{g}\rangle\) and \(|\mathrm{e}\rangle\) remains uncoupled.  The $\ket{\mathrm{s}} \leftrightarrow \ket{\mathrm{e}}$, transition is resonant with the cavity. Therefore, the coupling constant  $g_\mathrm{g} = 0$ whereas $g_\mathrm{s} \equiv g \gg \kappa, \gamma$.  We assume that the light-matter interaction operates in the strong coupling regime. i.e., $\frac{g^2}{\kappa\gamma}>1$.

\begin{center}
  \begin{figure}
\captionsetup{justification=myjust, singlelinecheck=false}

  \includegraphics[width=\columnwidth]{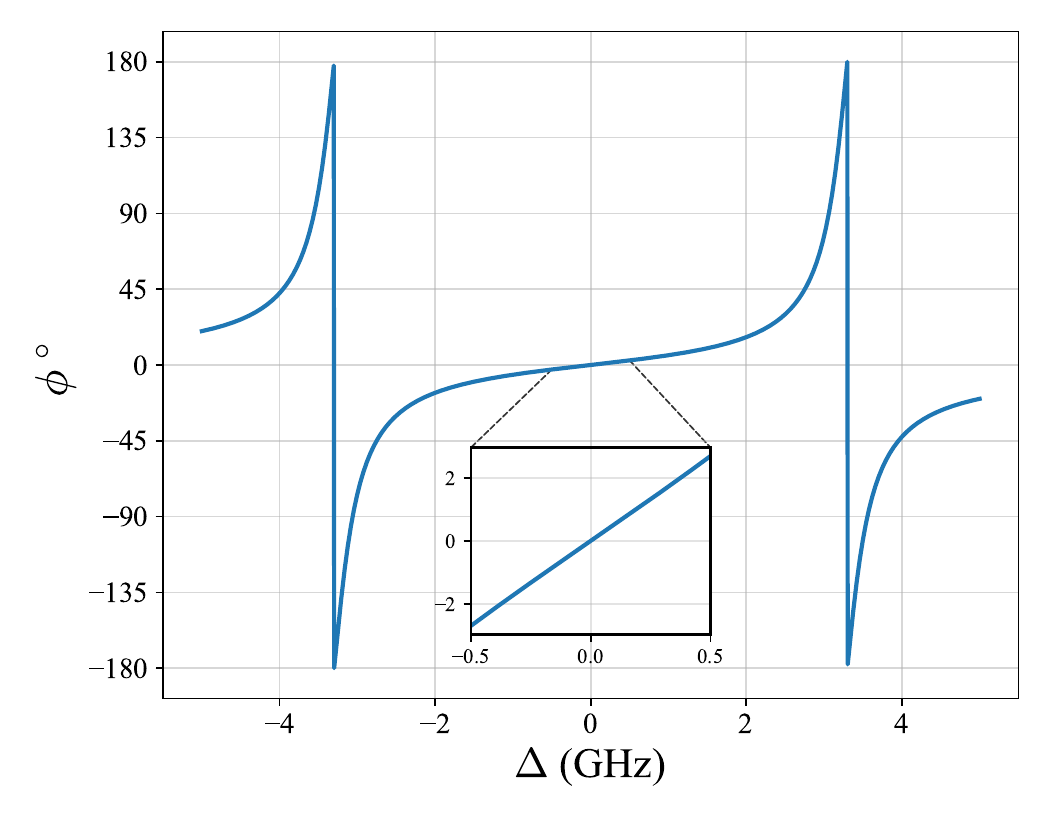}
\caption{Reflection phase, $\phi$ as a function of emitter–cavity detuning $\Delta$, computed from Eq.~\eqref{eq:reflection_general}, for a strongly coupled system ($g = 3.31~\mathrm{GHz}$, $\kappa = 1~\mathrm{GHz}$, $\gamma = 140~\mathrm{MHz}$, $\kappa_{\rm loss} = 10~\mathrm{MHz}$). The main panel shows the full dispersive response, while the inset zooms into the region near resonance ($\Delta \approx 0$)
For ideal cat-state generation, the reflected pulse should acquire a uniform phase. However, a finite-duration pulse has a finite bandwidth, causing different frequency components to experience different phases, which reduces fidelity. This distortion is suppressed when the pulse bandwidth is small compared to the phase variation scale, requiring $\frac{g^2}{\kappa}\sigma \gg 1$, where $\sigma$ is the pulse duration. This provides a key condition for high-fidelity cat-state generation.
}
    \label{fig:PhaseVsDetuning}
\end{figure}
\end{center}

While Eq.~\eqref{eq:reflection_general} for the reflection coefficient $r(\omega_p)$ suffices for computing mean output fields, it does
not provide direct access to the full quantum state of the scattered radiation,
particularly when the input field occupies a well-defined temporal mode. This could be understood from Fig.~\ref{fig:PhaseVsDetuning}. In this figure, we plot the phase acquired by a monochromatic light upon reflection from the strongly coupled quantum dot-cavity system as a function of detuning $\Delta_c = \Delta_0 = \Delta$ for the case of strong coupling.

For a perfect cat-state generation, the phase acquired should be zero. However, if we have a narrow temporal pulse, then the spectrum of the pulse will be broad. That would cause certain frequency components to acquire non-zero phase, resulting in lower fidelity of the generated cat states Fig.~\ref{fig:PhaseVsDetuning}. In order to obtain a high fidelity cat-state, we need to satisfy
\begin{align}
\label{eq:pw_f_con}
  \frac{g^2}{\kappa} \sigma \gg 1.
\end{align}
where $\sigma$ is the pulse width. This condition ensures that the photon interacts with the cavity–emitter
system long enough for the reflection to impart a nearly uniform phase across
the entire pulse. Thus, Eq.~\eqref{eq:pw_f_con} directly ties the system's effective
interaction strength to the pulse duration, providing a simple guideline for
high-fidelity cat-state generation. 

We present the results from our numerical simulations in order to assess the quality of the cat-states and find the optimum parameters.

\subsection{QD-cavity interaction for Cat-State Generation}
\label{subsec:cat_generation_simulation}

Using the formalism described in Appendix~\ref{sec:io_theory}, we perform numerical simulations to study the interaction between a traveling optical
pulse and a strongly coupled QD-cavity system using the cascaded
input-output formalism. The incoming optical field is generated by a (virtual) input
cavity that emits a Gaussian temporal mode of width $\sigma$. The pulse is prepared
in a displaced squeezed state
\begin{equation}
\ket{\psi_{\mathrm{in}}} = \ket{\alpha, r},
\end{equation}
with displacement amplitude $\alpha = \sqrt{\pi}$ and squeezing parameter 
$r = 0.98$ (see Sec.~\ref{subsec:squeezing_Q0} and Sec.~\ref{subsec:telep_num}).

The input cavity is coupled to the system cavity with decay rate $\kappa$. The system cavity houses a $\Lambda$-type quantum emitter, shown in Fig.~\ref{fig:atom-cavity}, 
which is initialized in an equal superposition of its two degenerate ground states,
\begin{equation}
\ket{\psi}_{\mathrm{QD}} = \frac{1}{\sqrt{2}}(\ket{\mathrm{g}} + \ket{\mathrm{s}}).
\end{equation}
The cavity couples strongly only to the $\ket{\mathrm{e}}\leftrightarrow\ket{\mathrm{s}}$ optical
transition, with coupling strength $g$, while the $\ket{\mathrm{g}}$ state remains effectively
decoupled. The system cavity is further coupled to a (virtual) output cavity, which
acts as a mode-selective filter and extracts only the Gaussian temporal mode of
interest \cite{Kiilerich2020}.

We simulate the full quantum dynamics of the combined system consisting of the input
cavity, system cavity, output cavity, and quantum emitter. The parameters are chosen
as $g = 3.31~\mathrm{GHz}$, $\kappa = 1~\mathrm{GHz}$, and
$\gamma = 140~\mathrm{MHz}$, corresponding to experimentally compatible values in
state-of-the-art quantum dot-cavity platforms
\cite{Najer2019}.
Our protocol requires the application of a Hadamard operation on the two ground states
of the emitter, followed by a projective measurement in the
$\{\ket{\mathrm{g}},\ket{\mathrm{s}}\}$ basis, which probabilistically yields even or odd Schrödinger
cat states of the output optical field.

\begin{figure*}[t]
  \centering
\captionsetup{justification=myjust, singlelinecheck=false}

  \begin{subfigure}[b]{0.9\columnwidth}
    \centering
    \includegraphics[width=\linewidth]{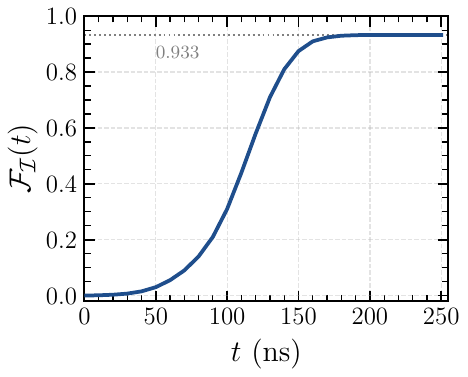}
    \caption{}
    \label{fig:fidelity}
  \end{subfigure}
  \hfill
  \begin{subfigure}[b]{\columnwidth}
    \centering
    \includegraphics[width=\linewidth]{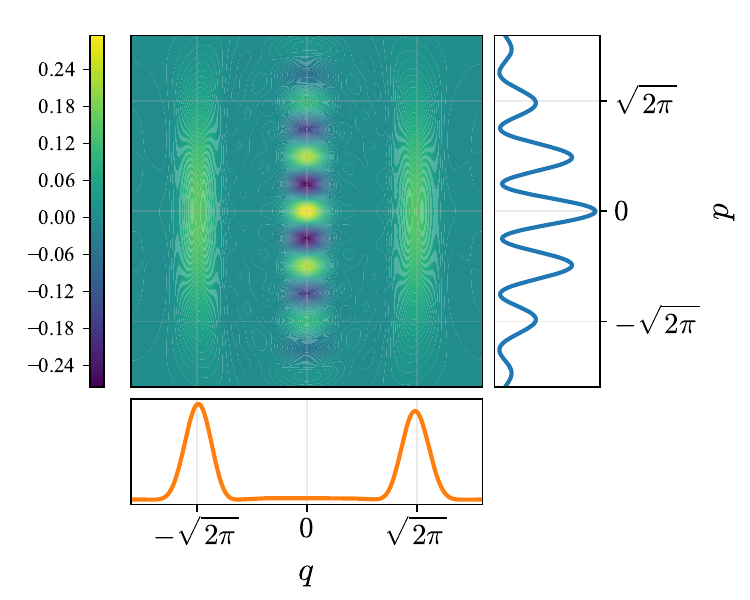}
    \caption{}
    \label{fig:Wig-gen}
  \end{subfigure}
  \caption{Fidelity and Wigner distribution of the cat state generated using a QD-cavity system.
  \textbf{(a)}~Time evolution of the fidelity $\mathcal{F_I}(t) = \langle R | \rho_{\mathrm{out}}(t) | R \rangle$ between the simulated joint state of the quantum emitter and the output virtual cavity and the ideal entangled reference state $|R\rangle = (|\mathrm{s}\rangle|\alpha,r\rangle + |\mathrm{g}\rangle|-\alpha,r\rangle)/\sqrt{2}$. The fidelity is near zero at early times because the output cavity is initially unpopulated (vacuum state); it rises as the pulse is transferred into the output mode and asymptotically saturates when all light has been collected.
  \textbf{(b)}~Wigner distribution of the conditional photonic state, $\rho_{\mathrm{cat}}$ obtained after applying a Hadamard gate and measuring the quantum dot spin. Conditioning on outcome $|\mathrm{s}\rangle$ yields the even squeezed cat state $|\mathrm{SC}_+(\alpha,r)\rangle$, while outcome $|\mathrm{g}\rangle$ yields the odd cat state $|\mathrm{SC}_-(\alpha,r)\rangle$; the distribution shown corresponds to the even cat state.
  Parameters:
  $g = 3.31~\mathrm{GHz}$, $\kappa = 1~\mathrm{GHz}$, $\gamma = 140~\mathrm{MHz}$,
  $\sigma = 25~\mathrm{ns}$, and $\kappa_{\mathrm{loss}} = 0$.
  The cat-state amplitude is $\alpha=\sqrt{\pi}$ and the squeezing parameter is
  $r=0.98$.}
  \label{fig:cat_results}
\end{figure*}

Ideally, prior to the application of the Hadamard operation and the quantum dot
measurement, the quantum emitter and the output cavity acquire the entangled
state
\begin{equation}
\ket{R} =
\frac{1}{\sqrt{2}}
\left(
\ket{\mathrm{s}}\ket{\alpha,r}
+
\ket{\mathrm{g}}\ket{-\alpha,r}
\right),
\label{eq:r_eq}
\end{equation}
which we use as the reference state for fidelity calculations. 

To quantify the quality of the generated entangled state, we define the
instantaneous fidelity
\begin{equation}
\mathcal{F_I}(t)
=
\bra{R}\,\rho_{\mathrm{out}}(t)\,\ket{R},
\label{eq:fidelity_def}
\end{equation}
where $\rho_{\mathrm{out}}(t) = \mathrm{Tr}_{\mathrm{in}}\left[\rho(t)\right]$
is the reduced density matrix of the joint system comprising the quantum emitter
and the output virtual cavity, obtained by tracing out the input cavity degrees
of freedom. 
In Fig.~\ref{fig:fidelity}, we plot $\mathcal{F_I}(t)$ as a function
of time. Initially, $\mathcal{F_I}(t) \approx 0$, since the
output virtual cavity is still in the vacuum state — no light has yet been
collected into the output mode, and the vacuum has zero overlap with the
displaced squeezed states $\ket{\pm\alpha, r}$ that comprise $\ket{R}$.
As time progresses and the pulse is gradually transferred from the physical
cavity into the output mode, the output cavity begins to populate and
$\mathcal{F_I}(t)$ rises accordingly. Once
all of the light couples into the output mode, the
fidelity asymptotically saturates, indicating the maximum achievable overlap
between the generated state and $\ket{R}$.

To obtain the final photonic state, a Hadamard gate is first applied to the
quantum dot spin in the $\{\ket{\mathrm{g}}, \ket{\mathrm{s}}\}$ basis,
rotating the entangled state $\ket{R}$ into
\begin{equation}
\mathcal{H}\ket{R} =
\frac{1}{2}\Big[
\ket{\mathrm{s}}\big(\ket{\alpha,r} + \ket{-\alpha,r}\big)
+
\ket{\mathrm{g}}\big(\ket{\alpha,r} - \ket{-\alpha,r}\big)
\Big].
\end{equation}
A subsequent projective measurement of the quantum dot spin then collapses
the optical field onto a definite cat state. Conditioning on outcome
$\ket{\mathrm{s}}$ yields the even squeezed cat state
$\ket{\mathrm{SC}_+(\alpha,r)} \propto \ket{\alpha,r} + \ket{-\alpha,r}$,
while conditioning on $\ket{\mathrm{g}}$ yields the odd cat state
$\ket{\mathrm{SC}_-(\alpha,r)} \propto \ket{\alpha,r} - \ket{-\alpha,r}$.
In practice, the output state is not the pure ideal cat state but rather
the reduced state of the output virtual cavity,
\begin{equation}
\rho_{\mathrm{cat}} =
\mathrm{Tr}_{\mathrm{QD}}\!\left[
\Pi_{\mathrm{s/g}}\,\rho_{\mathrm{out}}(t_f)\,\Pi_{\mathrm{s/g}}
\right],
\end{equation}
where $\Pi_{\mathrm{s/g}} = \ket{\mathrm{s}}\bra{\mathrm{s}}$ or
$\ket{\mathrm{g}}\bra{\mathrm{g}}$ is the projector corresponding to the
measurement outcome, and $\rho_{\mathrm{out}}(t_f)$ is the joint
emitter--output-cavity density matrix at the end of the interaction,
i.e.\ the state whose fidelity is plotted in Fig.~\ref{fig:fidelity} at
its saturated value. In Fig.~\ref{fig:Wig-gen}, we show the Wigner
distribution of $\rho_{\mathrm{cat}}$ for the outcome $\ket{\mathrm{s}}$,
corresponding to the even cat state $\ket{\mathrm{SC}_+(\alpha,r)}$.

\begin{figure}[h]
\captionsetup{justification=myjust, singlelinecheck=false}

  \includegraphics[width=0.35\textwidth]{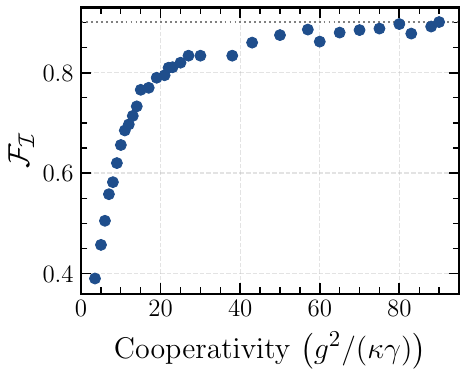}
  \caption{
Fidelity versus cooperativity $g^2/(\kappa \gamma)$, revealing that the loss-induced limitation can be overcome by increasing the interaction strength relative to dissipation. The clear, monotonic rise in fidelity demonstrates that the $\sim 93\%$ ceiling is not fundamental, but technological. This establishes cooperativity as a central parameter controlling coherence in the protocol. Parameters: $\gamma = 140~\text{MHz}$, $\sigma = 25~\text{ns}$, $\kappa_{\text{loss}} = 1\%$, $\alpha = \sqrt{\pi}$, $r = 0.98$.
  }
  \label{fig:coop-fid}
\end{figure}

Although the Wigner distribution of the final state closely resembles that of an
ideal cat state (see Fig.~\ref{fig:Wig-cat}), the fidelity saturates at approximately
$93.3\%$. This reduction is primarily due to the finite free-space decay rate
$\gamma$ of the emitter, which couples photons into undesired modes and thereby
reduces coherence. Increasing the ratio $g^2/(\kappa\gamma)$ or cooperativity would suppress this effect and improve the achievable fidelity. This is seen in Fig.~\ref{fig:coop-fid} wherein we vary the cooperativity by sweeping the $g$ and $\kappa$ while keeping $\gamma$ fixed. This directly probes the competition between coherent light–matter interaction and free-space loss.

\begin{figure*}[t]
\captionsetup{justification=myjust, singlelinecheck=false}

  \begin{subfigure}[b]{0.6\columnwidth}
    \includegraphics[width=1\linewidth]{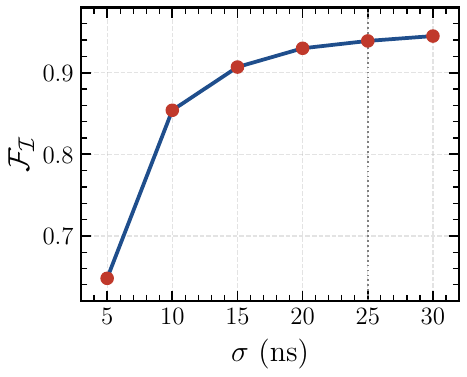}
    \caption{}
    \label{fig:FidVsPulse}
  \end{subfigure}
  \begin{subfigure}[b]{0.6\columnwidth}
    \includegraphics[width=1\linewidth]{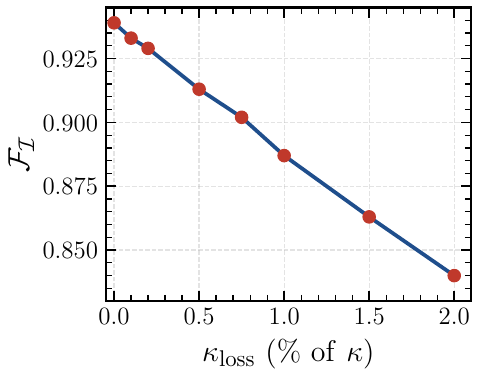}
    \caption{}
    \label{fig:FidVskappal}
  \end{subfigure}
  \begin{subfigure}[b]{0.6\columnwidth}
    \includegraphics[width=\linewidth]{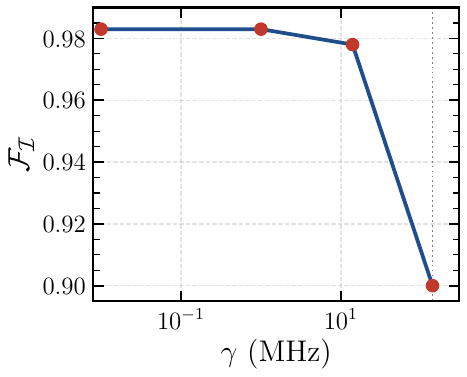}
    \caption{}
    \label{fig:FidVsgamma}
  \end{subfigure}
  
  \caption{Effect of pulse width and losses on cat-state fidelity. Subfig.~\ref{fig:FidVsPulse} shows the fidelity as a function of pulse width $\sigma$ for $\kappa_{\mathrm{loss}}=0$ and $\gamma = 140~\mathrm{MHz}$. Subfig.~\ref{fig:FidVskappal} shows the fidelity as a function of $\kappa_{\mathrm{loss}}$ for $\sigma=25~\mathrm{ns}$ and $\gamma = 140~\mathrm{MHz}$. Subfig.~\ref{fig:FidVsgamma} shows the fidelity as a function of $\gamma$ for $\sigma=25~\mathrm{ns}$ and $\kappa_{\mathrm{loss}}$. The dot-cavity coupling rate, $g =3.31~\mathrm{GHz}$, and $\kappa = 1~\mathrm{GHz}$.}

  \label{fig:sweepresults}

\end{figure*}

Additional factors influencing the quality of the generated cat states include the
pulse width $\sigma$ and the intrinsic cavity loss rate $\kappa_{\mathrm{loss}}$.
In Fig.~\ref{fig:sweepresults}, we explore the dependence of the fidelity on these
parameters. Fig.~\ref{fig:FidVsPulse} shows the fidelity of the final optical state
as a function of the pulse width $\sigma$ for $\kappa_{\mathrm{loss}}=0$. For the
chosen values of $g$, $\kappa$, and $\gamma$, short pulses
($\sigma \lesssim 25~\mathrm{ns}$) yield poor fidelity due to spectral mismatch with
the cavity, whereas for 
$\sigma > 25~\mathrm{ns}$ the fidelity asymptotically
saturates near $93\%$.

Figure~\ref{fig:FidVskappal} shows the effect of cavity loss on the cat-state
fidelity for fixed pulse width, $\sigma = 25~\mathrm{ns}$. As $\kappa_{\mathrm{loss}}$ increases,
the fidelity decreases approximately linearly, reflecting the direct loss of quantum
coherence. A similar effect is observed for ~\ref{fig:FidVsgamma}, which highlights the effect of $\gamma$. The frequency is represented in logarithmic scale for this plot for visualization. The fidelity trend would be approximately linear if the frequency was also in linear scale.

\subsection{Generating qunaught states}
\label{sec:gen_q0}

\begin{figure}[h]
\captionsetup{justification=myjust, singlelinecheck=false}

  \includegraphics[width=0.5\textwidth]{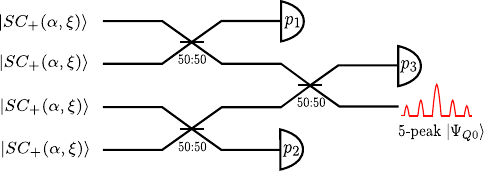}
\caption{Two-round cat-breeding protocol for generating the five-peak
qunaught state $|Q_0\rangle$. In Round~1, two independent breeding runs each interfere a pair of identical squeezed cat states
$|\mathrm{SC}_+(\alpha, r)\rangle$ on a balanced beam splitter.
Homodyne measurements ($p_1$ and $p_2$) on one of their respective output mode post-selects
a three-peaked intermediate grid state from the other. The two resulting
intermediate states are identical by construction and are fed as inputs
into Round~2, where a third beam splitter followed by a homodyne
measurement ($p_3$) produces the five-peaked output state. Accepting
homodyne outcomes within a high-fidelity window rather than strict
post-selection on a single value substantially increases the overall
success probability.}  \label{Fig:Cat-Breeding}
\end{figure}

As shown from Sec.~\ref{subsec:squeezing_Q0} and Sec.~\ref{subsec:telep_num}, the five-peak $\ket{Q_0}$ state with $8.5$ dB squeezing ($r=0.98$) is the optimum choice for our protocol. We can achieve this by two successive rounds of the cat-breeding protocol, with an important structural distinction between the two rounds. In particular, the second round of cat breeding requires \emph{two identical intermediate states} produced from the first round. These two intermediate states are then interfered and post-selected in the second round to yield a state that more closely approximates the desired $Q_{0}$ state.

\begin{figure*}[t]
  \centering

  \begin{subfigure}[b]{\columnwidth}
    \includegraphics[width=1\textwidth]{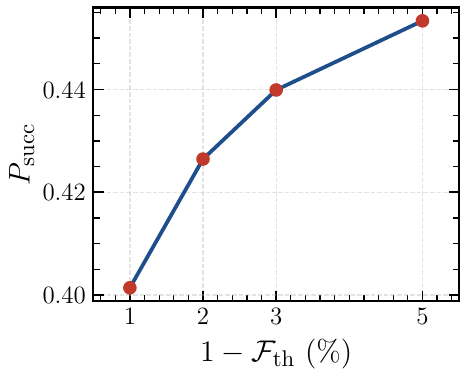}
    \caption{}
    \label{fig:1fcb}
  \end{subfigure}
  \hfill
  \begin{subfigure}[b]{\columnwidth}
    \includegraphics[width=1\textwidth]{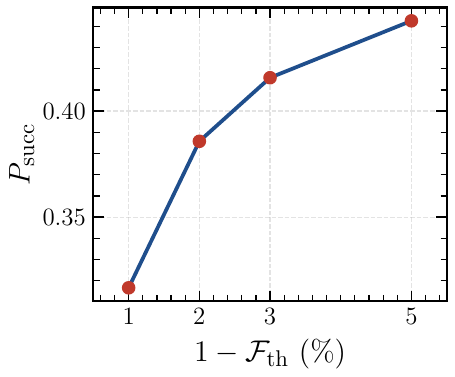}
    \caption{}
    \label{fig:2fcb}
  \end{subfigure}

  \caption{The probability of $|Q_0\rangle$ generation swept against the threshold for acceptable fidelity. 
  Subfig.~\ref{fig:1fcb} shows the one-fold cat-breeding success rate for a given fidelity threshold. 
  Subfig.~\ref{fig:2fcb} shows the same for two-fold cat breeding.}

  \label{fig:two-fold-succ-rate}
\end{figure*}

    In the first round of cat breeding, two identical squeezed cat states are interfered on a balanced beam splitter and conditioned on homodyne measurement outcomes $p$ belonging to a selected acceptance set (see  Sec.~\ref{sec:improving_success} and Fig.~\ref{fig:3pfid}. By allowing a small infidelity tolerance of approximately $1\%$, i.e.,
\begin{equation}
    \mathcal{F}^{(1)}(p) \geq 0.99,
\end{equation}
and by accepting all homodyne outcomes within the corresponding high-fidelity regions, the success probability for generating a single intermediate bred-cat state can be increased to approximately (see Fig.~\ref{fig:1fcb}
\begin{equation}
    P_{\mathrm{succ}}^{(1)} \simeq 0.4
\end{equation}
This intermediate state already exhibits a partial grid-like structure in phase space, but does not yet correspond to the $Q_{0}$ state.

Crucially, the second round of cat breeding requires \emph{two such identical intermediate states} as inputs. Since each intermediate state is generated probabilistically, the probability of successfully preparing the required pair is
\begin{equation}
    P_{\mathrm{pair}} = \left( P_{\mathrm{succ}}^{(1)} \right)^2 \simeq 0.16
\end{equation}

These two intermediate states are then interfered on a second balanced beam splitter, and one of the output modes is subjected to a homodyne measurement. Conditioning on the measurement outcome $p$ in this second round produces a  five peak state which occasionally resembles the $\ket{Q_0}$ state.

 As in the first round, the fidelity of the conditional output state in the second round depends sensitively on the homodyne outcome $p$. We define the second-round fidelity with respect to the target $Q_{0}$ state as
\begin{equation}
    \mathcal{F}^{(2)}(p) = \langle Q_{0} | \rho^{(2)}(p) | Q_{0} \rangle .
\end{equation}
By accepting all measurement outcomes satisfying
\begin{equation}
    \mathcal{F}^{(2)}(p) \geq \mathcal{F}_{\mathrm{th}},
\end{equation}
with a chosen threshold $\mathcal{F}_{\mathrm{th}}$, the success probability of the second breeding step remains substantial. The total probability of successfully generating the $Q_{0}$ state is therefore given by
\begin{equation}
    P_{\mathrm{succ}}^{(Q_{0})}
    = P_{\mathrm{pair}} \times P_{\mathrm{succ}}^{(2)},
\end{equation}
where $P_{\mathrm{succ}}^{(2)}$ denotes the conditional success probability of the second round under the same relaxed post-selection criterion. The success rate for the second round of the cat-breeding for infidelity of $<1\%$ is about $0.3$ (see ~\ref{fig:2fcb}). This results in a breeding success rate of a $\ket{Q_0}$ state to be $P_{\rm succ}^{Q_0} \approx (0.4)^2\times 0.3 \approx 4.8\%.$ 

While the QD-cavity source deterministically emits a cat state on
every attempt, the final projective spin measurement selects between
the even and odd cat, $|SC_+(\alpha,r)\rangle$ or $|SC_-(\alpha,r)\rangle$,
with equal probability. Since the breeding protocol requires two
\emph{identical}-parity cats as inputs, only a $1/2$ of
the generation attempts yield a usable pair. The
overall end-to-end success probability for the full protocol is - 

\begin{equation}
P_{\rm succ}^{Q_0} \times \frac{1}{2}
\approx 4.8\% \times 0.5 \approx 2.4\%.
\end{equation}

\section{Comparison with Recent Experimental GKP Architectures}
\label{sec:comparison_gkp_cluster}

Recent years have witnessed remarkable experimental progress in the generation of
approximate GKP states and CV cluster
states, particularly in photonic platforms aimed at fault-tolerant quantum
computation. In this section, we compare two representative state-of-the-art
experimental schemes with the protocol developed in this work, highlighting
complementary strengths, limitations, and industry relevance.

Large-scale photonic architectures based on multiplexing and modular design have
demonstrated the feasibility of generating GKP states and cluster states
using squeezed-light sources and adaptive measurements. But the generation rates still remain as low as 10s of Hz, despite stringent loss budgets and substantial optical overhead.
\cite{AghaeeRad2025, Larsen2025}. In parallel, optical platforms have realized a GKP generation rate of 10 Hz using bulk optics and high-level squeezing \cite{Konno2024}. However, their reliance on bulk components and limited modularity poses challenges for large-scale integration.

By contrast, the protocol presented here focuses on the efficient generation of
high-fidelity $|Q_0\rangle$ (square-lattice GKP) resource states using deterministic
light-matter interactions in a cavity-QED setting. Rather than relying on massive
multiplexing or ultra-high squeezing, our approach exploits state-dependent phase
shifts and optimized cat-breeding protocols to achieve high fidelities with
significantly relaxed post-selection criteria. This positions our scheme as a compact,
hybrid alternative that is naturally compatible with chip-scale photonics and optical
quantum networks.

Overall, while existing photonic platforms emphasize either large-scale multiplexing
or high squeezing to achieve fault-tolerant GKP resources, our approach provides a
complementary pathway centered on deterministic emitter–photon interactions. This
opens the door to hybrid architectures in which compact, high-quality non-Gaussian
resource states generated in cavity-QED systems are interfaced with larger photonic quantum processors and networks.
\label{subsec:comparison_results}

\section{Conclusion and Outlook}
\label{sec:conclusion}
In this work, we have presented a detailed and experimentally motivated framework for the generation of finite-energy \emph{qunaught} states in an optical setting. Our results demonstrate that high-quality approximate qunaught states can be generated using a combination of deterministic light-matter interactions and an optimized post-selection strategy, addressing limitations of existing optical approaches.

The proposed protocol consists of three main stages. First, we showed that spin-dependent reflection from a $\Lambda$-type quantum dot embedded in a strongly coupled optical cavity enables the deterministic generation of high-fidelity squeezed Schr\"odinger cat states. This stage eliminates the reliance on probabilistic cat-state preparation methods commonly employed in optical platforms and leverages recent advances in solid-state cavity quantum electrodynamics\cite{Hastrup2022}. Second, these squeezed cat states are converted into grid-like states through a multiround cat-breeding protocol based on balanced beam-splitter interference and homodyne detection. Third, and crucially, we demonstrated that the success probability of this breeding process can be significantly enhanced by relaxing the conventional post-selection strategy\cite{Pizzimenti25}. By accepting a structured set of homodyne measurement outcomes that maintain a high overlap with the target $\lvert Q_0 \rangle$ state, rather than post-selecting on a single measurement value, the protocol achieves a favorable trade-off between state fidelity and heralding probability.

Using numerical simulations, we quantified this trade-off for experimentally relevant parameters and showed that approximate qunaught states with high fidelity can be generated at success probabilities that exceed those of standard cat-breeding schemes by orders of magnitude. We further analyzed the effects of realistic imperfections, including finite pulse duration, and cavity photon loss. Our results identify parameter regimes consistent with current or near-term quantum dot--cavity experiments in which these imperfections do not qualitatively limit performance, thereby supporting the experimental feasibility of the approach.

Placing these results in the context of the current experimental landscape, it is worth noting that the most advanced demonstrations of grid-state preparation and error correction have thus far been realized in microwave-frequency and trapped-ion platforms. While these systems offer strong intrinsic nonlinearities and deterministic control, their operating frequencies and cryogenic requirements pose challenges for large-scale networking. Optical implementations, in contrast, are compatible with room-temperature operation, fiber-based quantum communication, and photonic integration, but have historically suffered from low success probabilities for non-Gaussian resource generation. By directly targeting the qunaught state using deterministic solid-state light-matter interactions combined with an enhanced post-selection strategy, the present work substantially mitigates this bottleneck and strengthens the case for optical grid-state architectures.

Taken together, our results establish a practically grounded roadmap for the generation of optical qunaught states using quantum dot-cavity systems. The protocol builds on well-established experimental components and utilizes a relaxed the post-selection criterion that makes the overall scheme significantly more practical. Looking ahead, this framework can be extended to higher-round breeding protocols, alternative lattice encodings, and direct integration with qunaught-based entanglement generation and fault-tolerant cluster-state constructions.

\bibliography{Ref_01}

\appendix

\section{Cascaded-System Formalism for Quantum Pulses}
\label{sec:io_theory}
\begin{figure}[t]
\captionsetup{justification=myjust, singlelinecheck=false}

  \centering
    \includegraphics[width=\columnwidth]{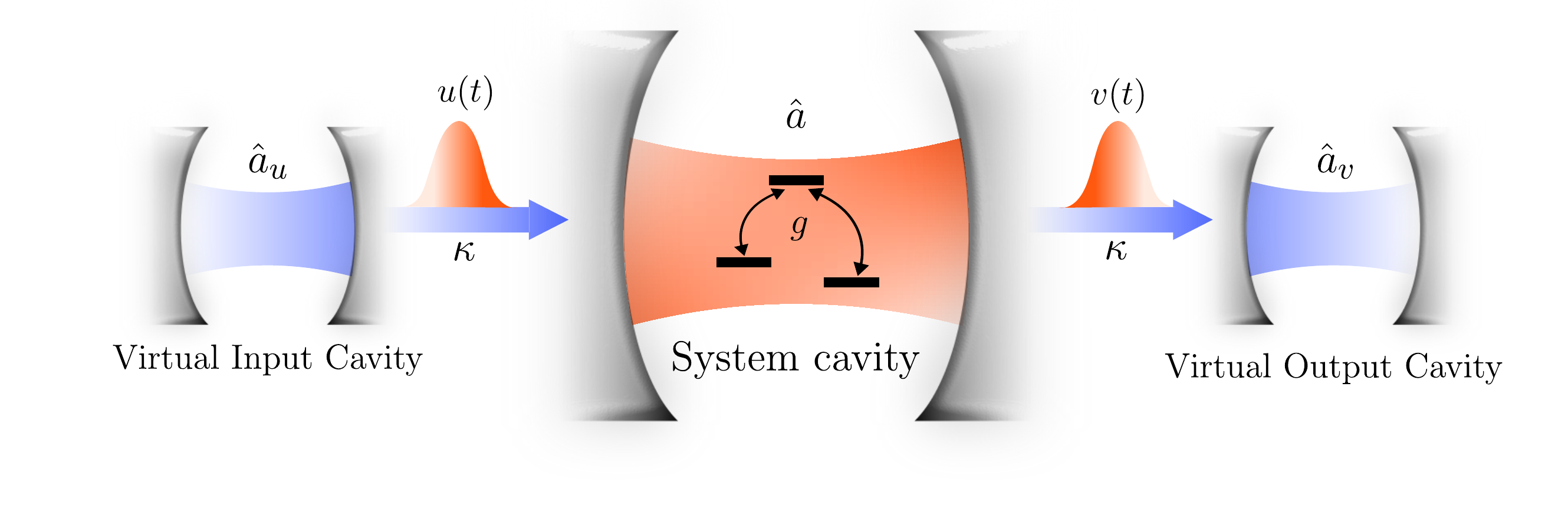}
  \caption{Cascaded three-cavity framework for simulating cat-state generation.
A displaced squeezed pulse $|\alpha, r\rangle$ in temporal mode $u(t)$ is
modelled by a virtual input cavity ($\hat{a}_u$), which drives the physical
system cavity ($\hat{a}$, decay rate $\kappa$) housing the $\Lambda$-type
quantum dot (coupling $g$, emission rate $\gamma$, intrinsic loss
$\kappa_{\rm loss}$). A virtual output cavity ($\hat{a}_v$) then
mode-selectively collects the scattered field into temporal mode $v(t)$,
giving direct access to the output state. The unidirectional cascaded
coupling ensures a consistent quantum treatment of the full pulse dynamics.}
  \label{fig:3_cav_diag}
\end{figure}

The reflection coefficient alone is sufficient for computing mean output fields, but it does not give us access to the quantum state of the scattered light when the input occupies a specific temporal mode - which is precisely what we need to evaluate cat-state fidelity. We therefore employ the cascaded-systems formalism, which provides a self-consistent quantum description of a shaped input pulse interacting with the cavity–emitter system and being collected into a defined output mode. This naturally accommodates finite pulse durations, cavity losses, and mode-selective detection, making it the right tool for a realistic simulation of the full protocol. To incorporate both the finite temporal width of the incoming pulse and a mode-selective description of the scattered field, we employ an extended
cascaded-systems formalism for quantum pulses that includes both an input
and an output virtual cavity \cite{Carmichael1993,Kiilerich2020,Kiilerich2019}.
A schematic of the three-cavity (input–system–output) setup is shown in Fig.~\ref{fig:3_cav_diag}.

In this approach, an incoming quantum pulse with normalized temporal mode
function $u(t)$ is modeled as a virtual input cavity with annihilation
operator $\hat{a}_u$. This construction provides a convenient way to generate a wavepacket with a prescribed temporal profile.

While the physical system emits radiation into a continuum of temporal modes,
the protocol of interest depends on the quantum state in a specific outgoing mode
$v(t)$. The virtual output cavity, with annihilation operator $\hat{a}_v$, acts as a
mode-selective receiver that collects the component of the field in this desired temporal mode. This allows us to map the multimode output field
onto a single bosonic degree of freedom, enabling direct computation of the
final quantum state and its fidelity.
Both virtual cavities are coupled unidirectionally to the physical system, ensuring a consistent cascaded description of the input–output dynamics.

The total Hamiltonian of the cascaded system \cite{Kiilerich2019, Kiilerich2020} is given by 
\begin{align}
\hat{H}
&=
\hat{H}_{\mathrm{JC}}
+
\frac{i\hbar}{2}
\Big[
\sqrt{\kappa}\, g_u(t)\, \hat{a}_u^\dagger \hat{a}
+
\sqrt{\kappa}\, g_v^*(t)\, \hat{a}^\dagger \hat{a}_v
\nonumber\\
&\quad
+g_u^*(t)g_v(t)\hat{a}_u^\dagger\hat{a}_v - \mathrm{h.c.}
\Big]
\label{eq:cascade_H_io}
\end{align}
where $\hat{a}$ is the annihilation operator of the physical cavity mode,
$\kappa$ denotes its decay rate, and $\hat{H}_{\mathrm{JC}}$ is the intrinsic
system Hamiltonian.
The dynamics are governed by a single time-dependent Lindblad operator,
\begin{equation}
\hat{L}_0(t)
=
\sqrt{\kappa}\,\hat{a}
+
g_u^*(t)\,\hat{a}_u
+
g_v(t)\,\hat{a}_v^\dagger,
\label{eq:L0_io}
\end{equation}
which ensures unidirectional flow of quantum information from the input
pulse cavity to the system and subsequently into the output pulse cavity.
The time-dependent coupling function $g_u(t)$ is chosen such that the virtual
input cavity emits the desired incoming temporal mode $u(t)$,
\begin{equation}
g_u(t)
=
\frac{u^*(t)}
{\sqrt{1 - \int_0^t |u(t')|^2\,dt'}}.
\label{eq:gu}
\end{equation}
Similarly, the coupling $g_v(t)$ is constructed to absorb the outgoing field
into the normalized temporal mode $v(t)$,
\begin{equation}
g_v(t)
=-
\frac{v(t)}
{\sqrt{\int_t^{\infty} |v(t')|^2\,dt'}}.
\label{eq:gv}
\end{equation}
The instantaneous output intensity is given by
\begin{equation}
I_{\mathrm{out}}(t)
=
\left\langle
\hat{L}_0^\dagger(t)\hat{L}_0(t)
\right\rangle,
\label{eq:Iout}
\end{equation}
and the first-order temporal correlation function by
\begin{equation}
g^{(1)}(t,t')
=
\left\langle
\hat{L}_0^\dagger(t)\hat{L}_0(t')
\right\rangle.
\label{eq:g1}
\end{equation}
Diagonalization of the kernel $g^{(1)}(t,t')$ yields an orthonormal basis of
output temporal modes and their corresponding occupation numbers, enabling
a complete mode-resolved characterization of the scattered quantum field.

\section{Effects of Non-Degenerate Ground States: Precession and Mitigation}
\label{subsec:nondegenerate_ground_states}
So far, we have considered the idealized situation in which the two ground states of
the $\Lambda$ system are degenerate. While this assumption is useful for building
intuition, it is generally not satisfied in realistic solid-state implementations.
In particular, non-degenerate ground states are often required in order to implement
single-qubit operations—such as Hadamard rotations—on the ground-state manifold.
Examples include Zeeman-split spin states in quantum dots or hyperfine-split states
in atomic systems \cite{Reiserer2015,Lodahl2015}.
The introduction of non-degeneracy, however, comes at the cost of coherent
precession and additional dephasing, both of which can degrade the quality of the generated Schrödinger cat states. In this subsection, we analyze how ground-state non-degeneracy affects cat-state generation and demonstrate its impact and mitigation. 

We incorporate this non-degeneracy ground state by adding a Zeeman term
\begin{equation}
\hat{H}_{\rm pre}
=
-\frac{1}{2}\hbar\omega_{gs}
\left(
|\mathrm{g}\rangle\langle \mathrm{g}|
-
|\mathrm{s}\rangle\langle \mathrm{s}|
\right),
\label{eq:HB}
\end{equation}
to $\hat{H}$~\eqref{eq:cascade_H_io} where $\hbar\omega_{gs}$ denotes the energy splitting between the states
$\ket{\mathrm{g}}$ and $\ket{\mathrm{s}}$. Physically, this splitting may arise from an external magnetic field or intrinsic anisotropies of the emitter.

\emph{Precession-induced phase accumulation:}
A direct consequence of the Hamiltonian~\eqref{eq:HB} is coherent precession of the
ground-state superposition which introduces a
time-dependent relative phase between $\ket{\mathrm{g}}$ and $\ket{\mathrm{s}}$, which is subsequently
mapped onto the optical field through the state-dependent cavity interaction.

\begin{figure}
\captionsetup{justification=myjust, singlelinecheck=false}

  \includegraphics[width=0.5\textwidth]{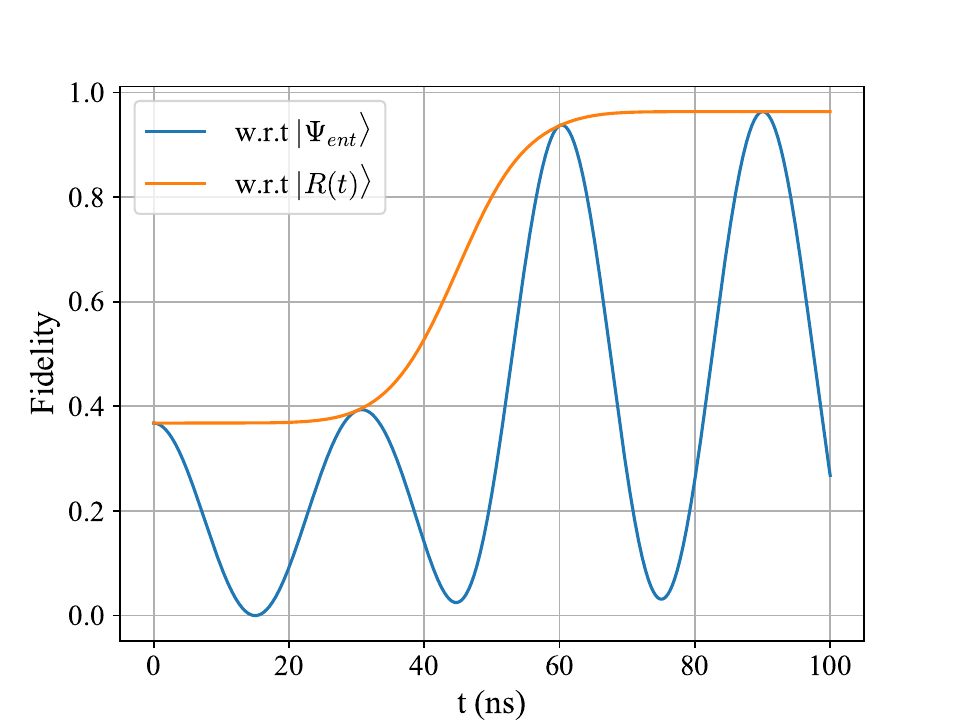}
  \caption{Fidelity of the cat state from non-degenerate $\Lambda$-system. The simulation used $\alpha=1$ and $r=0$, to ensure the 2 cavity approximation. Other system parameters: $g = 3.31~\mathrm{GHz}$, and $\kappa = 1~\mathrm{GHz}$, $\gamma=140~\mathrm{MHz}$, $\sigma=25~\mathrm{ns}$, $\kappa_{\mathrm{loss}}=0.01\kappa$.}
  \label{fig:precession_fidelity}
\end{figure}

To confirm that the observed fidelity oscillations originate from coherent
precession, we compare the output state to a time-dependent reference cat state
\begin{equation}
\ket{R(t)} = \ket{\alpha,r} + e^{i\omega_{gs} t}\ket{-\alpha,r},
\end{equation}
which explicitly includes the phase accumulated due to ground-state splitting.
Fig.~\ref{fig:precession_fidelity} shows the fidelity between the generated optical
state and $|R\rangle$ in ~\eqref{eq:r_eq} as a function of time. The pronounced oscillatory
behavior of the fidelity indicates a periodic relative phase accumulation between
the two coherent-state lobes of the cat state. This behavior is a clear signature of
ground-state precession. Fig~\ref{fig:precession_fidelity} also shows the fidelity of the output state with respect to
$\ket{R(t)}$, demonstrating that the dominant effect of non-degeneracy is indeed a
coherent, predictable phase rotation rather than irreversible decoherence. 

\end{document}